\documentclass[aps,preprint,11pt,nofootinbib]{revtex4-1}
\pdfoutput=1
\usepackage{amsmath, bm}
\usepackage{graphicx}
\usepackage{float}
\usepackage{xspace}
\newcommand{\UPMNS}{$U_{\text{PMNS}}$\xspace}

\begin{document}

\title{Nonzero $\theta_{13}$ in $SO(3) \rightarrow A_4$ lepton models}
\author{Yuval Grossman}
\email{yg73@cornell.edu}
\affiliation{Laboratory for Elementary-Particle Physics, Cornell University, Ithaca, N.Y.}
\author{and Wee Hao Ng}
\email{wn68@cornell.edu}
\affiliation{Laboratory for Elementary-Particle Physics, Cornell University, Ithaca, N.Y.}


\begin{abstract}
The simplest neutrino mass models based on $A_4$ symmetry predict $\theta_{13} = 0$ at tree level, a value that contradicts recent data. We study models that arise from the spontaneous breaking of an $SO(3)$ symmetry to its $A_4$ subgroup, and find that such models can naturally accommodate a nonzero $\theta_{13}$ at tree level. Standard Model charged leptons mix with additional heavy ones to generate a $\theta_{13}$ that scales with the ratio of the $A_4$-breaking to $SO(3)$-breaking scales. A suitable choice of energy scales hence allows one to reproduce the correct lepton mixing angles. We also consider an alternative approach where we modify the alignment of flavons associated with the charged lepton masses, and find that the effects on $\theta_{13}$ are enhanced by a factor that scales as $m_\tau/m_\mu$.
\end{abstract}

\maketitle

%
%

\section{Introduction}

For a long time, the Pontecorvo-Maki-Nakagawa-Sakata (PMNS) matrix \UPMNS \cite{ref:MNS1, ref:MNS2} was believed to be consistent with the tri-bimaximal mixing matrix \cite{ref:tribi}:
\begin{equation}
U_{\text{TBM}} = \left( \begin{array}{ccc} \sqrt{\frac{2}{3}} & \frac{1}{\sqrt{3}} & 0 \\ -\frac{1}{\sqrt{6}} & \frac{1}{\sqrt{3}} & \frac{1}{\sqrt{2}} \\ \frac{1}{\sqrt{6}} & -\frac{1}{\sqrt{3}} & \frac{1}{\sqrt{2}} \end{array} \right).
\end{equation}
The pattern exhibited by the tri-bimaximal mixing matrix seemed to suggest some underlying symmetry in the lepton sector, thus motivating the development of lepton models based on discrete flavor symmetries. One class of models was based on the discrete group $A_4$ \cite{ref:A4_1, ref:A4_2, ref:A4_3, ref:A4_4, ref:A4_5, ref:A4_6, ref:A4_7, ref:A4_8, ref:A4_9, ref:A4_10, ref:A4_11, ref:A4_12, ref:A4_13, ref:A4_14, ref:A4_15, ref:A4_16, ref:A4_17, ref:A4_18, ref:A4_19, ref:A4_20, ref:A4_21, ref:A4_22, ref:A4_23, ref:A4_24, ref:A4_25, ref:A4_26, ref:A4_27, ref:A4_28, ref:A4_29, ref:A4_31, ref:A4_32, ref:A4_33, ref:A4_34, ref:A4_35, ref:A4_36, ref:A4_37, ref:A4_38, ref:A4_39, ref:A4_40, ref:A4_41, ref:A4_42, ref:A4_43, ref:A4_44, ref:A4_45} (see \cite{ref:A4rev_1, ref:A4rev_2, ref:A4rev_3, ref:A4rev_4} for reviews). The most basic implementation of these $A_4$ models comprise the SM leptons and Higgs, right-handed singlet neutrinos, as well as two scalar flavons $\phi$ and $\phi'$. These fields are assigned into representations of $A_4$, where the flavons, in particular, are three-dimensional representations. The Lagrangian is invariant under $A_4$, but this symmetry is spontaneously broken when the flavons acquire VEVs, thus generating mass terms for the leptons. To reproduce the tri-bimaximal mixing at tree level, the lepton mass matrices have to take specific forms, which imply specific alignments $\langle \phi \rangle = (v,v,v)$ and $\langle \phi' \rangle = (v',0,0)$ for the flavon VEVs. Such alignments may be explained by various UV completions based on supersymmetry or extra dimensions \cite{ref:alignment1, ref:alignment2, ref:alignment3, ref:alignment4, ref:alignment5, ref:alignment6, ref:alignment7, ref:alignment8}.

The recent discovery of nonzero $\theta_{13}$ by the Daya Bay~\cite{ref:Daya} and RENO~\cite{ref:RENO} experiments have thrown the tri-bimaximal mixing pattern into question. The current experimental status of the elements of \UPMNS is shown in Table~\ref{tab:current-status}. The best fit value of $|\sin(\theta_{13})|$ for both hierarchies is approximately 0.16, significantly different from zero. This can be interpreted in two different ways, the first being that there is really no symmetry behind the lepton mixing (anarchy), and the second that there is a large modification to tri-bimaximal mixing in the underlying symmetry models.

\begin{table}
\renewcommand{\tabcolsep}{8pt}
\centering
\begin{tabular} {| c | c | c | c | c |}
\hline
Parameter & Best fit & $1\sigma$ range & $3\sigma$ range\\
\hline
$\sin^2(\theta_{12})$ (NH, IH) & $0.307$ & $0.291-0.325$ & $0.259-0.359$\\
\hline
$\sin^2(\theta_{13})$ (NH) & $0.0243$ & $0.0216-0.0266$ & $0.0169-0.0313$\\
$\sin^2(\theta_{13})$ (IH) & $0.0242$ & $0.0219-0.0267$ & $0.0171-0.0315$\\
\hline
$\sin^2(\theta_{23})$ (NH) & $0.386$ & $0.365-0.410$ & $0.331-0.637$\\
$\sin^2(\theta_{23})$ (IH) & $0.392$ & $0.370-0.431$ & $0.335-0.663$\\
\hline
$\delta$ (NH) & $1.08 \pi$ & $0.77\pi - 1.36\pi$ & $-$\\
$\delta$ (IH) & $1.09 \pi$ & $0.83\pi - 1.47\pi$ & $-$\\
\hline
\end{tabular}
\caption{Current experimental status of the mixing angles in \UPMNS \cite{ref:expt}. NH and IH stands for normal and inverted hierarchy respectively.}
\label{tab:current-status}
\end{table}

In accordance with the second viewpoint, various ideas have been proposed to modify the $A_4$ models to reproduce a nonzero $\theta_{13}$. One way is to consider higher dimension operators, which introduce correction terms to the mass matrices of relative size given by $v/\Lambda$ and $v'/\Lambda$
(where $\Lambda$ is the cutoff scale)~\cite{ref:alignment8, ref:nlo1, ref:A4rev_3}. Another approach is to extend the $A_4$ model to include more flavons that contribute to the lepton mass matrices~\cite{ref:moreflavons11, ref:moreflavons12, ref:moreflavons13, ref:moreflavons2, ref:moreflavons18}. Yet another avenue is to introduce perturbations in the flavon sector that modify their vacuum alignments and hence the form of the lepton mass matrices~\cite{ref:misalign1, ref:misalign2, ref:moreflavons14, ref:misalign3}. Radiative corrections as a way to generate nonzero $\theta_{13}$ have also been considered in~\cite{ref:radiative10, ref:radiative9, ref:radiative11, ref:radiative12, ref:radiative13}.

In this paper, we focus on a specific class of models \cite{ref:other-so(3)1, ref:berger, ref:other-so(3)2} that can be regarded as UV completions of certain $A_4$ models. These UV models are invariant under a continuous symmetry group, for example $SO(3)$, of which $A_4$ is a subgroup. This symmetry is spontaneously broken to $A_4$ by certain flavons that acquire a specific pattern of VEVs, generating $A_4$ models as effective low energy theories. In light of the recent measurements, it is worth investigating how a nonzero $\theta_{13}$ can be accommodated in such models.

An interesting observation is that that such models actually already predict $\theta_{13}$ to be nonzero even with the usual vacuum alignment. $SO(3)$-based models in general require additional heavy charged leptons to complete the $SO(3)$ representations the Standard Model (SM) charged leptons belong to. While the mixing between SM and heavy charged leptons is very small, it is enough to modify the pattern of the light charged lepton mass matrix, which breaks the tri-bimaximal mixing pattern. The idea of modifying the charged-lepton mass matrix to obtain a nonzero $\theta_{13}$ is certainly not new; however, seldom has the context been that of mixing between SM and heavy charged leptons. This will be the main focus of our work, using a model motivated by \cite{ref:berger} as an illustration. An interesting result is that the size of $\theta_{13}$ scales with the ratio of $A_4$-breaking to $SO(3)$-breaking scales. In other words, $\theta_{13}$ may reflect certain features of the UV physics, rather than simply arising from some arbitrary coefficients.

A second way to obtain a nonzero $\theta_{13}$, with a clear parallel in typical $A_4$ models, is to allow the VEVs to deviate from the usual alignment that reproduces $U_{\text{TBM}}$. An interesting feature of this approach is the presence of an enhancement factor that scales as $m_\tau/m_\mu$ should the flavons involved be those associated with the charged lepton masses. In other words, a small angular deviation of these flavons from the usual alignment can give rise to a much larger $\theta_{13}$.

This paper is organized as follows. In Section~\ref{sect:review} we provide an overview of the $SO(3) \rightarrow A_4$ model of~\cite{ref:berger}. In Section~\ref{sect:first-method} we present our main results, where we demonstrate that mixing of the SM charged leptons with heavy ones give rise to a nonzero tree-level $\theta_{13}$, the size of which is related to the ratio of scales.  In Section~\ref{sect:second-method}, we discuss the second approach of modifying the flavon alignment associated with charged leptons and demonstrate the existence of the enhancement factor. We summarize our work in Section~\ref{sect:conclusion}. Details of the model and the numerical simulations are given in the appendix.

%
%

\section{Review of the $SO(3) \rightarrow A_4$ model} \label{sect:review}

\subsection{Field content}

We review a lepton model motivated by \cite{ref:berger} where a larger continuous flavour symmetry is spontaneously broken to the $A_4$ subgroup. The symmetries of this model are the electroweak gauge symmetry $SU(2)_L \times U(1)_Y$ as well as a global $SO(3)_F$. The fields and their representations are summarized in Table~\ref{tab:fields}.

For the lepton sector, the three SM left-handed $SU(2)_L$ doublets $\psi_l$ form a $\bm{3}$ under $SO(3)_F$. Among the three SM charged right-handed $SU(2)_L$ singlets, $\psi_e$ is a $\bm{1}$, while the other two have been subsumed into a $\bm{5}$ denoted by $\psi_m$. In doing so, we now have three extra charged right-handed $SU(2)_L$ singlets from $\psi_m$, which we give large masses to by introducing a charged left-handed $SU(2)_L$ singlet $\psi_f$ that form a $\bm{3}$. Finally we introduce three right-handed neutrinos $\psi_n$ which form a $\bm{3}$.

In the scalar sector, we have the SM Higgs, $H$, which is a singlet of the flavor group, and four flavons $\phi$, $\phi'$, $\phi_5$, and $T$ which are $\bm{3}$, $\bm{3}$, $\bm{5}$ and $\bm{7}$ respectively. The flavon $T$ is responsible for the $SO(3)_F\to A_4$ breaking, and is required to be at least a $\bm{7}$ since that is the smallest representation of $SO(3)$ that can have an $A_4$-invariant VEV. While $\phi$ and $\phi'$ can be identified with the usual flavons in the minimal $A_4$ model, the extra flavon $\phi_5$ is required here to prevent the muon and tau from becoming degenerate. This degeneracy is due to the right-handed muon and tau being part of the same $SO(3)_F$ multiplet and hence sharing the same Yukawa coupling with $\phi$. An extra flavon $\phi_5$ in a different $SO(3)_F$ representation from $\phi$ is needed to lift this degeneracy. (We note that this degeneracy is actually also lifted by the block-diagonalization process to be discussed later, but the resulting mass differences are in general too small.) 

\begin{table}
\renewcommand{\tabcolsep}{8pt}
\centering
\begin{tabular} {| c | c | c | c |}
\hline
Field & $SU(2)_L$ & $U(1)_Y$ & $SO(3)_F$\\
\hline
$\psi_l$ & $\bm{2}$ & $-\frac{1}{2}$ & $\bm{3}$\\
$\psi_f$ & $\bm{1}$ & $-1$ & $\bm{3}$\\
\hline
$\psi_e$ & $\bm{1}$ & $-1$ & $\bm{1}$\\
$\psi_m$ & $\bm{1}$ & $-1$ & $\bm{5}$\\
$\psi_n$ & $\bm{1}$ & $0$ & $\bm{3}$\\
\hline
$H$ & $\bm{2}$ & $\frac{1}{2}$ & $\bm{1}$\\
$\phi$ & $\bm{1}$ & $0$ & $\bm{3}$\\
$\phi'$ & $\bm{1}$ & $0$ & $\bm{3}$\\
$\phi_5$ & $\bm{1}$ & $0$ & $\bm{5}$\\
$T$ & $\bm{1}$ & $0$ & $\bm{7}$\\
\hline
\end{tabular}
\caption{Matter fields and the representations they transform as under the gauge symmetry $SU(2)_L \times U(1)_Y$ and global symmetry $SO(3)_F$. We have partitioned the fields into left-handed leptons, right-handed leptons and scalars.}
\label{tab:fields}
\end{table}

\subsection{Lagrangian} \label{sect:lagrangian}

We now assume the following terms in the Lagrangian for the charged leptons and neutrinos:
\begin{align}
\mathcal{L}_e =& -y_e \overline{\psi_l}^a \frac{H}{\Lambda} \phi^a \psi_e -y_m \overline{\psi_l}^a \frac{H}{\Lambda} \phi^b \psi_m^{ab} - y^T_m \overline{\psi_l}^a \frac{H}{\Lambda} T^{abc} \psi_m^{bc} - y^5_m \epsilon^{abc} \overline{\psi_l}^a \frac{H}{\Lambda} \phi_5^{bd} \psi_m^{cd} \label{eqn:charged-lepton-lagrangian} \\
& -y'_e \overline{\psi_f}^a \phi^a \psi_e -y'_m \overline{\psi_f}^a \phi^b \psi^{ab}_m -y^{T\prime}_m \overline{\psi_f}^a T^{abc} \psi_m^{bc} - y^{5\prime}_m \epsilon^{abc} \overline{\psi_f}^a \phi_5^{bd} \psi_m^{cd} + \text{h.c.}, \nonumber \\
\mathcal{L}_\nu =& -M \overline{ \psi^{\text{c}}_n}^a \psi^a_n - \frac{x_\nu}{\Lambda} \overline{\psi^{\text{c}}_n}^a \psi^b_n \phi^{\prime c} T^{abc} - y_\nu \overline{\psi_l}^aH^{\text{c}} \psi^a_n + \text{h.c.}, \label{eqn:neutrino-lagrangian}
\end{align}
where $a$, $b$, and $c$ are $SO(3)_F$ indices running from $1$ to $3$, and $\Lambda$ is the cutoff scale of the model. This Lagrangian is not renormalizable and includes certain dimension-five operators. These are required to give masses to the light charged leptons, and to lift the degeneracy of the light neutrinos.

There are other terms in the Lagrangian involving only the scalars, of which we will just focus on the renormalizable self-interactions of the flavon $T$:
\begin{equation}
V(T) = -\frac{\mu^2}{2}T^{abc}T^{abc}+\frac{\lambda}{4}(T^{abc}T^{abc})^2+cT^{abc}T^{bcd}T^{def}T^{efa}.
\end{equation}
It is shown in \cite{ref:berger} that conditions on $\lambda$ and $c$ exist such that $V(T)$ has an $A_4$ invariant minimum, which breaks $SO(3)_F$ into its $A_4$ subgroup. We then end up with an effective non-minimal $A_4$ model, with three more pairs of left- and right-handed charged leptons, and one more flavon $\phi_5$.

Before proceeding further, we acknowledge two issues with the lepton Lagrangian. First, this is not the most general Lagrangian consistent with the gauge and global symmetries. \cite{ref:berger} proposed an auxiliary $Z_2$ symmetry to forbid the excluded terms, but a careful check shows that it does not work. Nonetheless, we have been able to identify modified models that can reproduce the same lepton mass matrices as this Lagrangian, the details of which are provided in Appendix~\ref{app:z-n-symm}. The second issue is that since the flavons can acquire VEVs of different scales, mass terms arising from higher-dimension operators need not be smaller than those from lower-dimension ones, especially if they contain multiple factors of the larger VEVs. Therefore, the errors associated with truncation of the Lagrangian can be significant. This issue will also be addressed in Appendix~\ref{app:z-n-symm}. For the rest of our work, we will neglect both issues and continue to work with the given Lagrangian to demonstrate the key ideas behind our approach.

\subsection{Lepton mass matrices and \UPMNS}

We assume that the flavons $\phi$, $\phi_5$ and $\phi'$ acquire VEVs with the following alignments:
\begin{equation}
\langle \phi \rangle = \left( \begin{array}{c} v \\ v \\ v \end{array} \right),
\qquad \langle \phi_5 \rangle = \left( \begin{array}{ccc} 0 & v_5 & v_5 \\ v_5 & 0 & v_5 \\ v_5 & v_5 & 0 \end{array} \right),
\qquad \langle \phi' \rangle = \left( \begin{array}{c} v' \\ 0 \\ 0 \end{array} \right).
\end{equation}
We also assume that the VEV of $T$ satisfies $v_T \gg v, v', v_5$, in accordance to the picture of $SO(3)_F$ broken to $A_4$. After electroweak symmetry breaking, the Higgs boson $H$ acquires a VEV $v_H = \left(246/\sqrt{2}\right)\, \text{GeV}$, and we obtain two $6 \times 6$ matrices: $M^{6 \times 6}_l$ for the charged-lepton Dirac masses and $M^{6 \times 6}_\nu$ for the neutrino Majorana masses.

In \cite{ref:berger}, the mixing between the SM and the new charged leptons were considered to be small and hence neglected. In that case, the leading mass matrix $M_l$ for the three light charged leptons is simply given by the upper-left $3 \times 3$ block of $M^{6 \times 6}_l$:
\begin{equation}
M_l = \frac{v_H}{\Lambda} \left( \begin{array}{ccc} y_e v & y_m v + y^5_m v_5(\omega^2-\omega) & y_m v + y^5_m v_5(\omega-\omega^2) \\
y_e v & [y_m v + y^5_m v_5(\omega^2-\omega)]\omega & [y_m v + y^5_m v_5(\omega-\omega^2)]\omega^2 \\
y_e v & [y_m v + y^5_m v_5(\omega^2-\omega)]\omega^2 & [y_m v + y^5_m v_5(\omega-\omega^2)]\omega \end{array} \right), \label{eqn:wrong-lepton-mass-matrix}
\end{equation}
where $\omega = e^{2\pi i/3}$. The charged lepton masses can be obtained by diagonalizing $M_l (M_l)^\dagger$ and taking the square root, and we obtain
\begin{equation}
m_e =\left\vert \sqrt{3} \frac{y_e v_H v}{\Lambda} \right\vert, \qquad m_\mu, m_\tau = \left\vert \sqrt{3} \frac{y_m v_H v}{\Lambda} \pm 3i \frac{y^5_m v_H v_5}{\Lambda} \right\vert. \label{eqn:mass-eigenvalues}
\end{equation}
We note that the Yukawas have to be fine-tuned to generate the correct charged lepton masses. Therefore, this model does not ameliorate the fine-tuning issue also present in minimal $A_4$ models\footnote{Certain $A_4$ models \cite{ref:A4rev_4, ref:alignment1} resolve the fine-tuning issue by relegating the electron mass to higher-dimensional operators, through the use of additional symmetries and flavons. While we do not show it here, an analogous approach can be adopted in our model to naturally suppress $y_e$. However, as we see later, this does not fully resolve the issue since subleading contributions from block diagonalization do not scale with $y_e$ in general.}.

The unitary transformation required to diagonalize $M_l (M_l)^\dagger$ is
\begin{equation}
U_l = \frac{1}{\sqrt{3}} \left( \begin{array}{ccc} 1 & 1 & 1 \\ 1 & \omega & \omega^2 \\ 1 & \omega^2 & \omega
\end{array} \right). \label{eqn:naive-ul}
\end{equation}
It is important to note that (\ref{eqn:naive-ul}) is a result of $M_l$ taking the form
\begin{equation}
M_{\text{aligned}} = \left( \begin{array}{ccc} a & b & c \\
a & b \,\omega & c \,\omega^2 \\
a & b \,\omega^2 & c \,\omega \end{array} \right), \label{eqn:m-aligned}
\end{equation}
where $a$, $b$ and $c$ are constants.

For the neutrino sector, the $6 \times 6$ Majorana mass matrix $M^{6 \times 6}_\nu$ can be block-diagonalized, and the resulting upper-left $3 \times 3$ block is identified with the mass matrix $M_\nu$ of the three light neutrinos 
\begin{equation}
M_\nu = -y^2_\nu v^2_H \left( \begin{array}{ccc} -\frac{1}{M} & 0 & 0 \\
0 & -\frac{M}{M^2 - x^2_\nu (v' v_T/\Lambda)^2} & \frac{x_\nu (v' v_T/\Lambda)}{M^2 - x^2_\nu (v' v_T/\Lambda)^2} \\[5pt]
0 & \frac{x_\nu (v' v_T/\Lambda)}{M^2 - x^2_\nu (v' v_T/\Lambda)^2} & -\frac{M}{M^2 - x^2_\nu (v' v_T/\Lambda)^2} 
\end{array} \right).
\end{equation}
Note that this is exactly the see-saw mechanism, as $M_\nu$ becomes very small if the Majorana mass parameters $M$ and $v' v_T / \Lambda$ for $\psi_n$ are much larger than $v_H$. The light neutrino masses can be obtained by diagonalizing $M_\nu$. We choose the following assignment for the mass eigenvalues:
\begin{equation}
m_1 = \left\vert \frac{y^2_\nu v^2_H}{M + x_\nu (v'v_T/\Lambda)} \right\vert, \quad m_2 = \left\vert \frac{y^2_\nu v^2_H}{M} \right\vert, \quad m_3 = \left\vert \frac{y^2_\nu v^2_H}{M - x_\nu (v'v_T/\Lambda)} \right\vert. \label{eqn:neutrino-mass}
\end{equation}
Such an assignment can accommodate both normal and inverted hierarchies, but the latter requires fine-tuning between the magnitude and phase of the combination $x_\nu v' v_T/(M \Lambda)$ \cite{ref:alignment8}. Therefore, for the rest of our work, we will only focus on normal hierarchy.

The unitary transformation required to diagonalize $M_\nu$ is:
\begin{equation}
U_\nu = i \left( \begin{array}{ccc} 0 & \frac{1}{\sqrt{2}} & \frac{1}{\sqrt{2}} \\ 1 & 0 & 0 \\ 0 & \frac{1}{\sqrt{2}} & -\frac{1}{\sqrt{2}}
\end{array} \right).
\end{equation}
The PMNS matrix is then given by:
\begin{equation}
U_{\text{PMNS}} = U_l(U_\nu)^\dagger = \left( \begin{array}{ccc} -i \sqrt{\frac{2}{3}} & -i\frac{1}{\sqrt{3}} & 0 \\ i\frac{1}{\sqrt{6}} & -i\frac{1}{\sqrt{3}} & \frac{1}{\sqrt{2}} \\ i\frac{1}{\sqrt{6}} & -i\frac{1}{\sqrt{3}} & -\frac{1}{\sqrt{2}} \end{array}\right), \label{eqn:naive-PMNS}
\end{equation}
which can be brought into the form $U_{\text{TBM}}$ by redefining the phases of $\nu_1$, $\nu_2$ and $\tau$. Note that the tri-bimaximal mixing pattern obtained above depends on $U_l$ taking the form Eq.~(\ref{eqn:naive-ul}). Any deviation of $M_l$ from $M_{\text{aligned}}$ would result in deviation of \UPMNS from $U_{\text{TBM}}$. We will exploit this fact in the next section in order to generate a nonzero $\theta_{13}$. We also note that Eq. (\ref{eqn:naive-PMNS}) omits certain nonunitary matrix factors which we show in Appendix \ref{app:nonunitary} to be negligible.

We now briefly mention the masses of the heavy leptons, which can be obtained from the corresponding full $6 \times 6$ mass matrices. The heavy charged lepton masses are typically of order $y^{T\prime}_m v_T$, and the heavy neutrino masses of order $M \sim x_\nu v'v_T/\Lambda$. (We will demonstrate shortly that $M$ and $x_\nu v'v_T/\Lambda$ are typically of the same scale.)

\subsection{Energy scales}\label{sect:energyscales}

We can use our results for the light lepton masses to obtain a rough picture of the energy scales in this model. We first consider the neutrino masses given in Eq.~(\ref{eqn:neutrino-mass}). The current experimental results are as follows \cite{ref:expt}:
\begin{align}
\delta m^2 &\equiv m_2^2 - m_1^2 = \, 7.54^{+0.26}_{-0.22}\times 10^{-5} \, \text{eV}^2,\\
\left| \Delta m^2 \right| &\equiv \left| m_3^2 - \frac{m_1^2+m_2^2}{2} \right| = \begin{array}{l} 2.43^{+0.06}_{-0.10} \, (\text{NH}) \\ 2.42^{+0.07}_{-0.11} \, (\text{IH}) \end{array} \times 10^{-3} \, \text{eV}^2.
\end{align}
Assuming normal hierarchy, we find that $M \sim x_\nu v'v_T/\Lambda \sim 10^{15} \vert y_\nu \vert \, \text{GeV}$, with neutrino masses $m_1 \sim 6 \, \text{meV}, m_2 \sim 10 \, \text{meV}$ and $m_3 \sim 50 \,\text{meV}$ independent of $y_\nu$. The fact that $M \sim x_\nu v'v_T/\Lambda$ is not surprising since we require large cancellations in $M - x_\nu v'v_T/\Lambda$ to ensure that $m_3 \gg m_1, m_2$, as implied by the experimental results. This gives rise to the following hierarchy of energy scales:
\begin{equation}
v_H \sim 100\, \text{GeV} \, \ll \, \frac{M}{x_\nu} \sim \frac{v'v_T}{\Lambda} \, \ll \, \{v, v_5, v'\} \, \ll \, v_T \, \ll \, \Lambda. \label{eqn:scales}
\end{equation}

The charged-lepton masses provide further constraints on the hierarchy. Since $m_\tau \sim 1 \, \text{GeV}$, Eq.~(\ref{eqn:mass-eigenvalues}) implies that $\{v, v_5, v' \}/\Lambda \gtrsim O(10^{-3})$ so that the associated Yukawas remain perturbatively small. This somewhat restricts the ratio of symmetry breaking scales $\epsilon \equiv \{v, v_5, v' \}/v_T$, if we do not want the scales $\{v, v_5, v' \}$, $v_T$ and $\Lambda$ to be too close to one another.

We now consider an example to illustrate the typical energy scales involved. Assuming all Yukawas to be $O(1)$, we find that $M \sim 10^{15} \, \text{GeV}$, $\{ v, v_5, v' \} \sim 10^{16} \, \text{GeV}$, $v_T \sim 10^{18} \, \text{GeV}$ and $\Lambda \sim 10^{19} \, \text{GeV}$. Other values can be obtained by varying the Yukawas, although this is limited by the requirement that the Yukawas remain perturbative.

%
%

\section{Effects of mixing in the charged lepton sector} \label{sect:first-method}

\subsection{Obtaining the light charged lepton mass-squared matrix}

The tri-bimaximal mixing pattern of this model is the result of the unitary matrix $U_l$ that diagonalizes $M_l (M_l)^\dagger$ taking the form Eq.~(\ref{eqn:naive-ul}). This in turn requires the light charged lepton mass matrix $M_l$ to take the form $M_{\text{aligned}}$ given in Eq.~(\ref{eqn:m-aligned}). As we shall see below, subleading corrections from block diagonalisation modifies the form of $M_l$ and hence lead to a nonzero $\theta_{13}$.

To obtain the exact form of $M_l (M_l)^\dagger$ in the general case, we start with the full $6 \times 6$ mass matrix $M^{6 \times 6}_l$ obtained from the Lagrangian Eq.~(\ref{eqn:charged-lepton-lagrangian}). We express $M^{6 \times 6}_l$ in terms of $3 \times 3$ matrices $A$, $B$, $C$ and $D$:
\begin{equation}
M^{6 \times 6}_l \equiv \left( \begin{array} {cc}
\frac{v_H}{\Lambda} A & \frac{v_H}{\Lambda} B\\
C & D \end{array} \right). \label{eqn:abcd}
\end{equation}
where
\begin{align}
A &= \left( \begin{array}{ccc}
y_e v & \left[ y_m v + y^5_m v_5(\omega^2-\omega) \right] & \left[ y_m v + y^5_m v_5(\omega-\omega^2) \right] \\
y_e v & \left[ y_m v + y^5_m v_5(\omega^2-\omega) \right] \omega & \left[ y_m v + y^5_m v_5(\omega-\omega^2) \right] \omega^2 \\
y_e v & \left[ y_m v + y^5_m v_5(\omega^2-\omega) \right] \omega^2 & \left[ y_m v + y^5_m v_5(\omega-\omega^2) \right] \omega \end{array} \right), \label{eqn:form-of-a}
\end{align}
\begin{align}
B &= \left( \begin{array}{ccc} y_m v + 2y^T_m v_T & y_m v + y^5_m v_5 & - y^5_m v_5\\
y_m v & 2 y^T_m v_T & y_m v\\
y_m v + y^5_m v_5 + y^T_m v_T & y_m v - y^5_m v_5 & y^T_m v_T \end{array} \right),
\end{align}
\begin{align}
C &= \left( \begin{array}{ccc} y'_e v & \left[ y'_m v + y^{5\prime}_m v_5(\omega^2-\omega) \right] & \left[ y'_m v + y^{5\prime}_m v_5(\omega-\omega^2) \right]\\
y'_e v & \left[ y'_m v + y^{5\prime}_m v_5(\omega^2-\omega) \right] \omega & \left[ y'_m v + y^{5\prime}_m v_5(\omega-\omega^2) \right] \omega^2 \\
y'_e v & \left[ y'_m v + y^{5\prime}_m v_5(\omega^2-\omega) \right] \omega^2 & \left[ y'_m v + y^{5\prime}_m v_5(\omega-\omega^2) \right] \omega \end{array} \right),\label{eqn:form-of-c}
\end{align}
\begin{align}
D &= \left( \begin{array}{ccc} y'_m v + 2y^{T\prime}_m v_T & y'_m v + y^{5\prime}_m v_5 & - y^{5\prime}_m v_5\\
y'_m v & 2 y^{T\prime}_m v_T & y'_m v\\
y'_m v + y^{5\prime}_m v_5 + y^{T\prime}_m v_T & y'_m v - y^{5\prime}_m v_5 & y^{T\prime}_m v_T \end{array} \right).
\end{align}
We then block-diagonalize $M^{6 \times 6}_l (M^{6 \times 6}_l)^\dagger$ and obtain $M_l (M_l)^\dagger$ from the upper-left $3 \times 3$ block:
\begin{equation}
M_l (M_l)^\dagger = \frac{v^2_H}{\Lambda^2} \left[ A A^\dagger + B B^\dagger - (A C^\dagger + B D^\dagger)(C C^\dagger + D D^\dagger)^{-1} (C A^\dagger + D B^\dagger) \right]. \label{eqn:full-mass-squared}
\end{equation}
We see that when we previously assumed $M_l$ to be given by Eq.~(\ref{eqn:wrong-lepton-mass-matrix}), we have kept only the leading term $v_H^2 AA^\dagger / \Lambda^2$. 

We now examine the effects of the other terms. We first note that $A$ and $C$ are both of the form $M_{\text{aligned}}$. We further define
\begin{equation}
E \equiv B - \frac{y^T_m}{y^{T\prime}_m} D,
\end{equation}
and the small parameter
\begin{equation}
\epsilon \equiv O(v/v_T) \sim O(v_5/v_T).
\end{equation}
Assuming all Yukawa couplings to be of the same order $y$, we find that the scales of $B$ and $D$ are of order $y v_T$, while those of $A$, $C$ and $E$ are of order $y v$ and hence $\epsilon$ smaller. (We quantify the scale of a matrix by the characteristic size of the eigenvalues). We can thus expand Eq.~(\ref{eqn:full-mass-squared}) in $\epsilon$. To the lowest nontrivial order we find that $M_l (M_l)^\dagger$ is factorizable with
\begin{equation}
M_l=\frac{v_H}{\Lambda} \left( A - BD^{-1}C \right) =
\frac{v_H}{\Lambda} \left( A - \frac{y^T_m}{y^{T\prime}_m} C - ED^{-1}C \right).
 \label{eqn:factorization2}
\end{equation}
Since both $A$ and $C$ are of the form $M_{\text{aligned}}$, so is any linear superposition of them, and thus the first correction term $-(y^T_m/y^{T\prime}_m) C$ does not give a nonzero $\theta_{13}$. The second correction term $-ED^{-1}C$, of order $\epsilon$, is what generate deviations of $M_l$ from $M_{\text{aligned}}$. This in turn suggests that the size of $\theta_{13}$ is also of order $\epsilon$. In other words, $\theta_{13}$ reflects the ratio of the $A_4$-breaking to $SO(3)_F$-breaking scales.

\subsection{Amplification from nearly-degenerate mass eigenvalues} \label{sect:amplification}

While the above analysis seems to suggest that $\theta_{13} \sim \epsilon$, there is actually a numerical factor that enhances the size of $\theta_{13}$. From perturbation theory, the mixing angle between two eigenvectors is given approximately by the ratio of the small perturbation mixing them to the difference between their eigenvalues. Our previous result of $\theta_{13} \sim \epsilon$ implicitly assumes that the difference between mass eigenvalues are of order $m_\tau$. However, in the actual charged-lepton mass spectrum, $m_e$ and $m_\mu$ are nearly degenerate relative to $m_\tau$, suggesting an enhancement in the mixing angle.

To illustrate this enhancement, it is useful to write $M_l (M_l)^\dagger$ in a different basis:
\begin{align}
U_l M_l (M_l)^\dagger (U_l)^\dagger &= \frac{v^2_H}{\Lambda^2} U_l \left( A - \frac{y^T_m}{y^{T\prime}_m} C \right) \left( A - \frac{y^T_m}{y^{T\prime}_m} C \right)^\dagger (U_l)^\dagger + \Delta\\
&= \left( \begin{array}{ccc} m_a^2 & 0 & 0 \\ 0 & m_b^2 & 0 \\ 0 & 0 & m_c^2 \end{array} \right) + \left( \begin{array}{ccc} \Delta_{11} & \Delta_{21}^* & \Delta_{31}^* \\ \Delta_{21} & \Delta_{22} & \Delta_{32}^* \\ \Delta_{31} & \Delta_{32} & \Delta_{33} \end{array} \right),
\end{align}
where $U_l$ takes the specific form given in Eq.~(\ref{eqn:naive-ul}) and
\begin{equation}
\Delta \equiv -\frac{v^2_H}{\Lambda^2} U_l \left[ \left( A - \frac{y^T_m}{y^{T\prime}_m} C \right) C^\dagger (D^\dagger)^{-1} E^\dagger + E D^{-1} C \left( A - \frac{y^T_m}{y^{T\prime}_m} C \right)^\dagger \right]U_l^\dagger, \label{eqn:delta}
\end{equation}
comes from the order $\epsilon$ corrections in Eq.~(\ref{eqn:factorization2}). In this basis, $\theta_{13}$ is determined by how much the perturbation $\Delta$ changes the zeroth-order eigenvector $(1,0,0)$. From perturbation theory, $\theta_{13}$ is roughly $\Delta_{21}/(m_b^2-m_a^2)$ or $\Delta_{31}/(m_c^2-m_a^2)$, whichever is larger.

Since the actual eigenvalues are given by the charged lepton masses, we assume the following sizes for the zeroth-order eigenvalues:
\begin{equation}
m_a^2 \lesssim m_\mu^2, \quad m_b^2 \sim m_\mu^2, \quad m_c^2 \sim m_\tau^2.
\end{equation}
The perturbation matrix $\Delta$ is of the scale $O(m_c^2 \epsilon)$, and so naively we might expect $\Delta_{21}/(m_b^2-m_a^2) \sim O(m_\tau^2 \epsilon/m^2_\mu)$ and $\Delta_{31}/(m_c^2-m_a^2) \sim O(\epsilon)$. This will imply that $\theta_{13}$ is enhanced compared to the naive expectation $\epsilon$ by $m_\tau^2/m_\mu^2$. However, as we show below, while indeed $\Delta_{31} \sim O(m_c^2 \epsilon)$, we instead find that $\Delta_{21} \sim O(m_b m_c \epsilon)$. This is due to $M_l (M_l)^\dagger$ being factorizable at a well-defined order in $\epsilon$, as we have demonstrated in Eq.~(\ref{eqn:factorization2}). As a result, $\Delta_{21}/(m_b^2-m_a^2) \sim O(m_\tau \epsilon/m_\mu)$, from which we conclude that 
\begin{equation}
\theta_{13} \sim O\left(\frac{m_\tau}{m_\mu}\epsilon\right). \label{eqn:main-result}
\end{equation}
Thus the size of $\theta_{13}$ is enhanced by a smaller factor $m_\tau/m_\mu$.

We now explain why the factorizability of $M_l (M_l)^\dagger$ at a well-defined order in $\epsilon$ sets the sizes of $\Delta_{21}$ and $\Delta_{31}$. In such a scenario, we expect $U_l M_l (U_l)^\dagger$ to be given by
\begin{equation}
U_l M_l (U_l)^\dagger = \left( \begin{array}{ccc} m_a & 0 & 0 \\
0 & m_b & 0 \\
0 & 0 & m_c
\end{array}
\right) + \epsilon m_c R
\end{equation}
where $R$ is an $O(1)$ matrix. Substituting this into $U_l M_l (M_l)^\dagger (U_l)^\dagger$, we find that
\begin{equation}
\Delta = \epsilon m_c \left( \begin{array}{ccc} m_a R_{11}^* + m_a R_{11} & m_a R_{21}^* + m_b R_{12} & m_a R_{31}^* + m_c R_{13} \\
m_b R_{12}^* + m_a R_{21} & m_b R_{22}^* + m_b R_{22} & m_b R_{32}^* + m_c R_{23} \\
m_c R_{13}^* + m_a R_{31} & m_c R_{23}^* + m_b R_{32} & m_c R_{33}^* + m_c R_{33}
\end{array}
\right)
\end{equation}
Therefore, we conclude that
\begin{align}
\Delta_{21} =& \epsilon m_c (m_b R_{12}^* + m_a R_{21}) \sim O(m_b m_c \epsilon),\\
\Delta_{31} =& \epsilon m_c (m_c R_{13}^* + m_a R_{31}) \sim O(m_c^2 \epsilon).
\end{align}
in agreement with our assertions above.

A few points to note: First, the analysis above breaks down when $\epsilon \gg m_\mu/m_\tau$, since this implies that the perturbation $\Delta_{21} \gg m_\mu^2$, in which case we also require the zeroth-order eigenvalues $m_a^2$, $m_b^2 \gg m_\mu^2$ so that large cancellations can occur to give two small eigenvalues $m_e^2$ and $m_\mu^2$. Second, as we shall see from the simulation results next section, there are various ``large'' $O(1)$ factors that we have not taken into account in our analysis, so the exact $\theta_{13}$ may be a few times smaller than the prediction above.

\subsection{Verifying the results via simulation} \label{sect:simulation1}

To verify the above results, we compute the exact tree-level \UPMNS for two large collections of random parameter sets $\mathcal{C}_1$ and $\mathcal{C}_2$. Details of how they are generated are provided in Appendix~\ref{app:random}. The mass eigenvalues for parameter sets in $\mathcal{C}_1$ are unconstrained, whereas those in $\mathcal{C}_2$ are required to have the correct charged-lepton mass ratios. We note that only a very small fraction of random parameter sets satisfy the conditions for $\mathcal{C}_2$, a consequence of the charged lepton mass fine-tuning.

\begin{figure}[t]
\centering
\includegraphics[width=.49\linewidth]{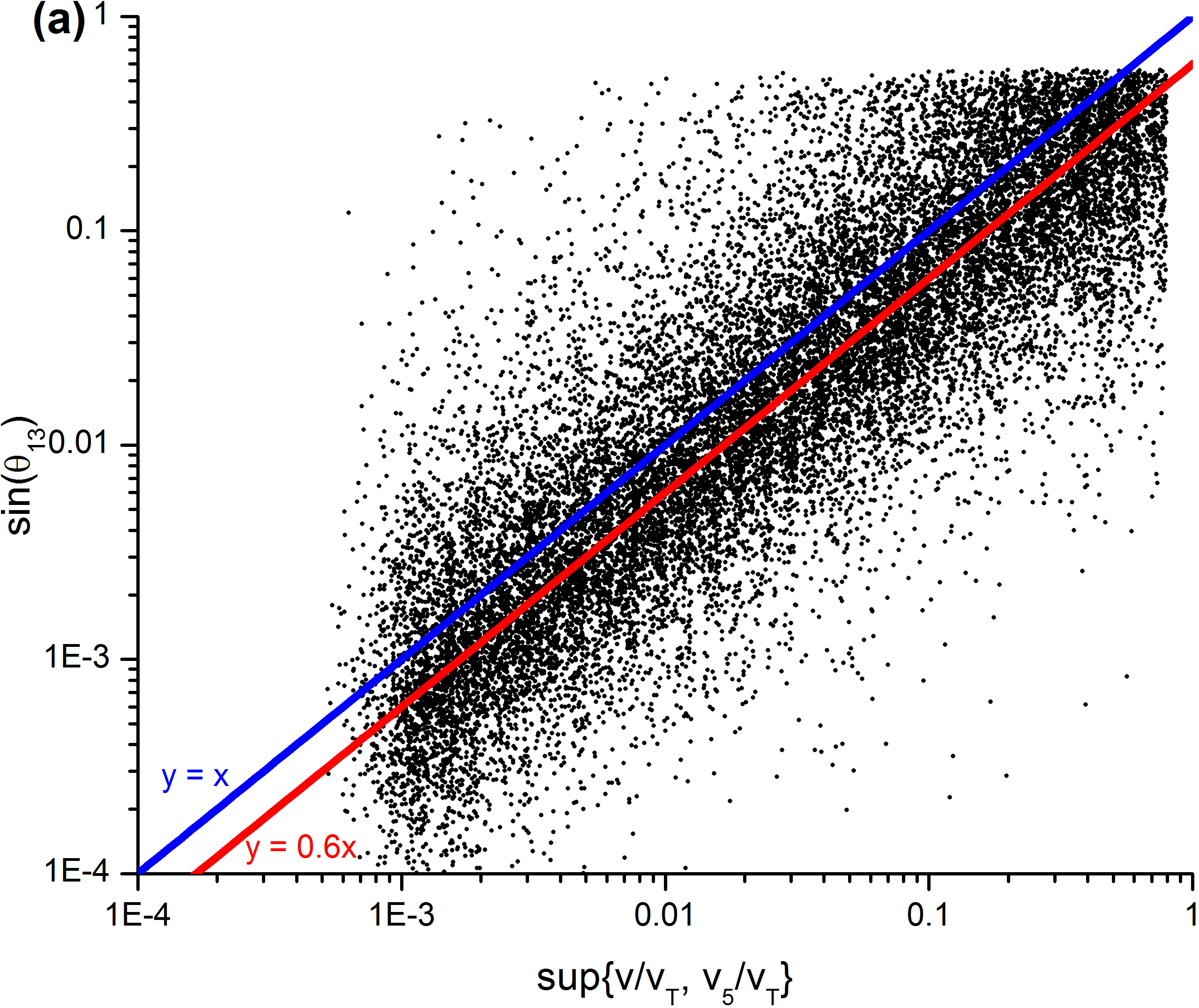}
\includegraphics[width=.49\linewidth]{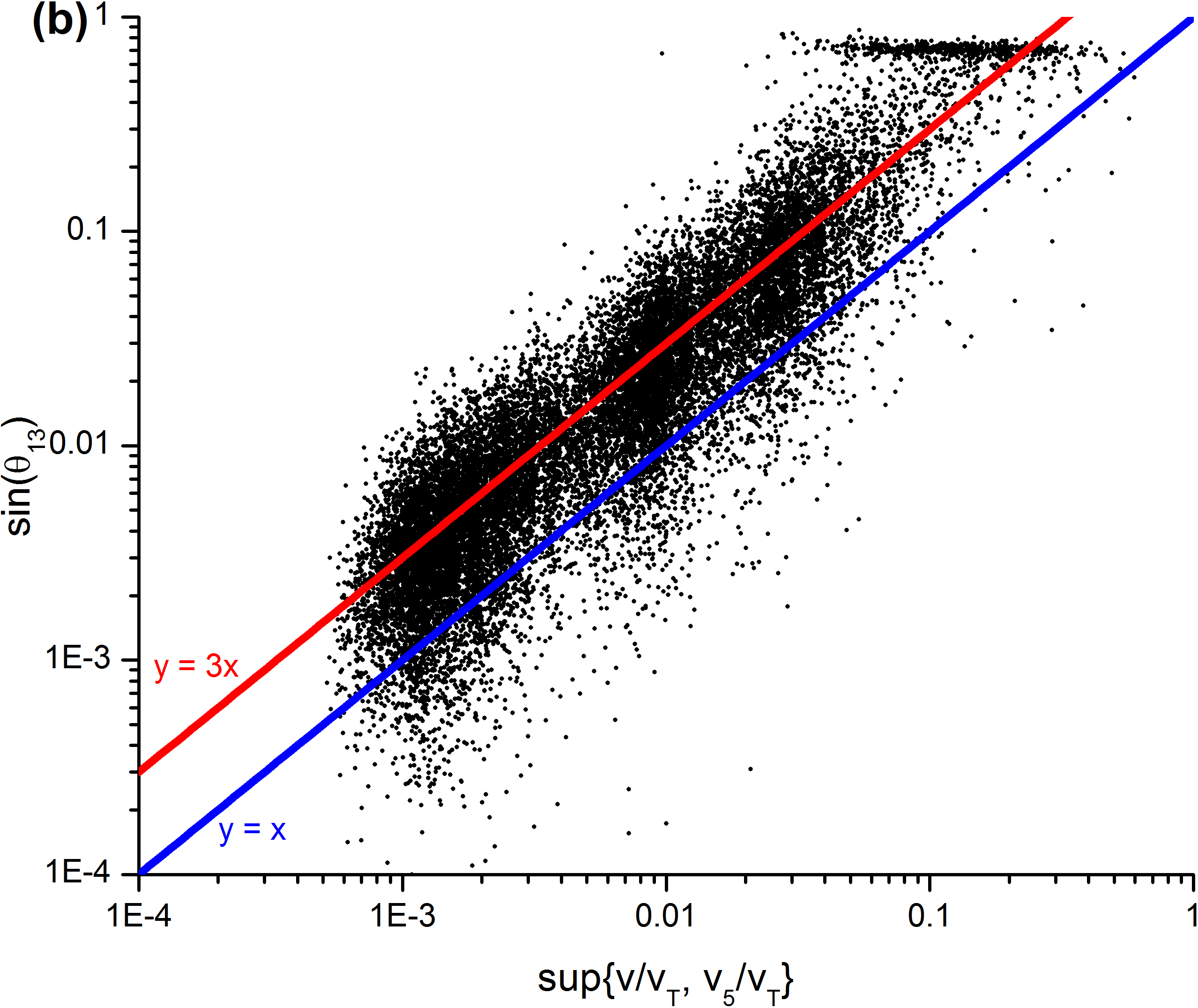}
\caption{\label{fig:first-method} Graphs of $\sin(\theta_{13})$ against $\sup\{{v/v_T}, {v_5/v_T} \}$ for two collections (a) $\mathcal{C}_1$ and (b) $\mathcal{C}_2$ of random parameter sets. Various lines have been included for reference. Both graphs demonstrate the linear dependence predicted in our analysis. Collection $\mathcal{C}_2$ has a charged lepton mass spectrum much closer to the actual hierarchy than $\mathcal{C}_1$, as a result of which the constant of proportionality is significantly enhanced by the amplification effect.}
\end{figure}

Figure~\ref{fig:first-method}(a) shows the value of $\sin(\theta_{13})$ against $\sup\{{v/v_T}, {v_5/v_T} \}$ for $\mathcal{C}_1$, which in general agrees with the expectation that $\theta_{13} \sim O(\epsilon)$. Since no conditions have been imposed on the mass ratios, all three mass eigenvalues are typically of the same order of magnitude and hence there is no significant amplification effect.

Figure~\ref{fig:first-method}(b) shows the value of $\sin(\theta_{13})$ against $\sup\{{v/v_T}, {v_5/v_T} \}$ for $\mathcal{C}_2$. While we still have $\theta_{13} \propto O(\epsilon)$, the proportionality constant is now about five times that of $\mathcal{C}_1$. Since the eigenvalues of $\mathcal{C}_2$ are now of the correct ratios, we attribute the larger proportionality constant to the amplification effect, although the amplification is smaller than our original prediction due to ``large'' $O(1)$ factors that we have neglected in our analysis. We hence conclude that the experimental best fit value of $|\sin(\theta_{13})| \simeq 0.16$ corresponds to the ratio of symmetry-breaking scales $\epsilon \sim 0.05$.

\subsection{Compatibility with experimental constraints}

We now discuss whether this model can satisfy the experimental constraints on lepton masses and mixing angles. We first consider lepton masses. As we have shown in Section~\ref{sect:energyscales}, the measured neutrino mass differences $\delta m^2$ and $|\Delta m^2|$ are certainly compatible with the model provided that $M \sim x_\nu v'v_T/\Lambda$. We have also demonstrated that the correct charged-lepton mass spectrum can be reproduced in Section~\ref{sect:simulation1}, although significant fine-tuning is required.

We now focus on the mixing angles. A major concern is that corrections that reproduce the measured $\theta_{13}$ might also affect the other mixing angles $\theta_{12}$ and $\theta_{23}$ to the extent that they no longer remain compatible with experimental observations. In particular, many models predict the same size of corrections to $\theta_{12}$ and $\theta_{13}$, in which case a large $\theta_{13}$ will imply a large correction to $\theta_{12}$.

\begin{figure}[t]
\centering
\includegraphics[width=.49\linewidth]{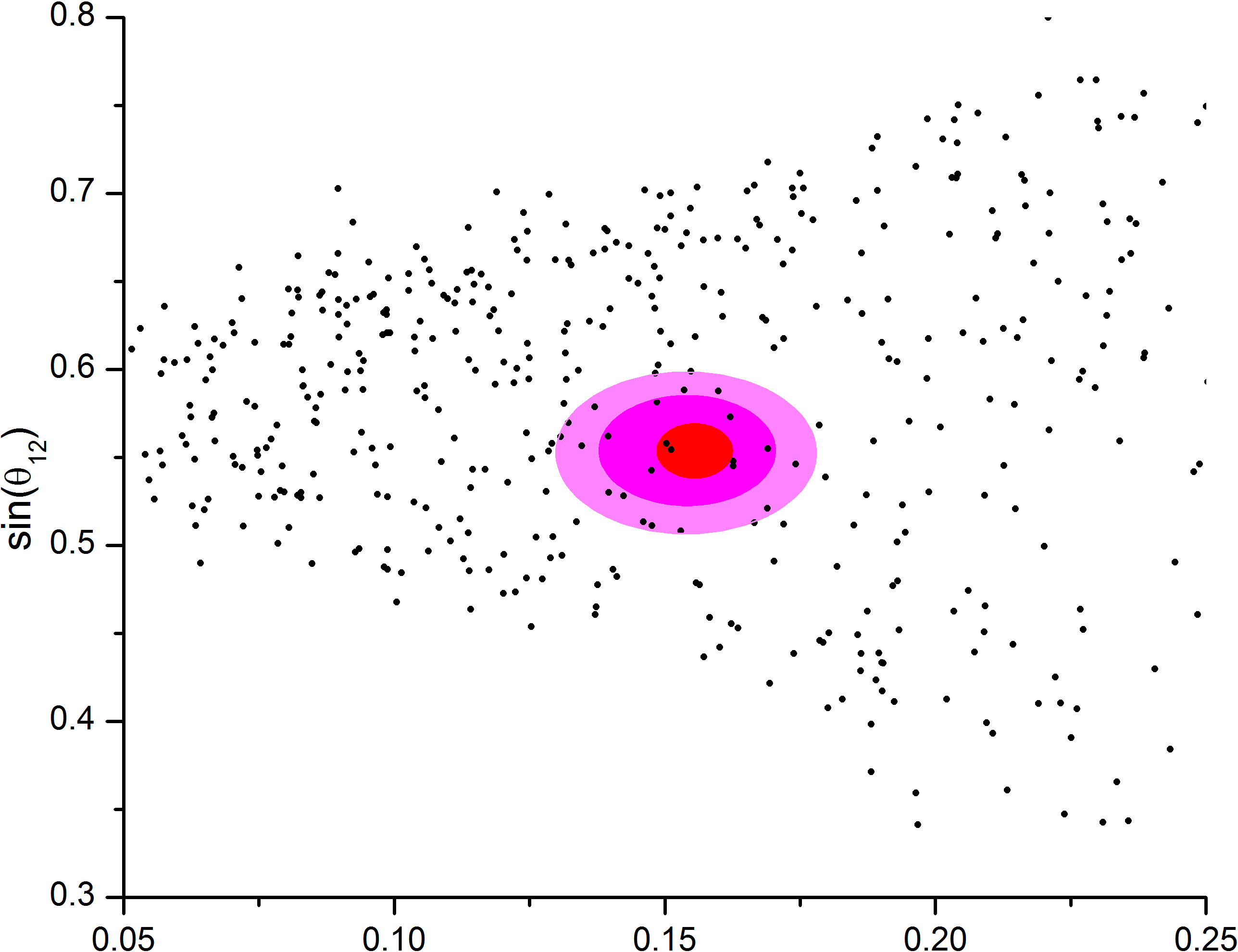}
\includegraphics[width=.49\linewidth]{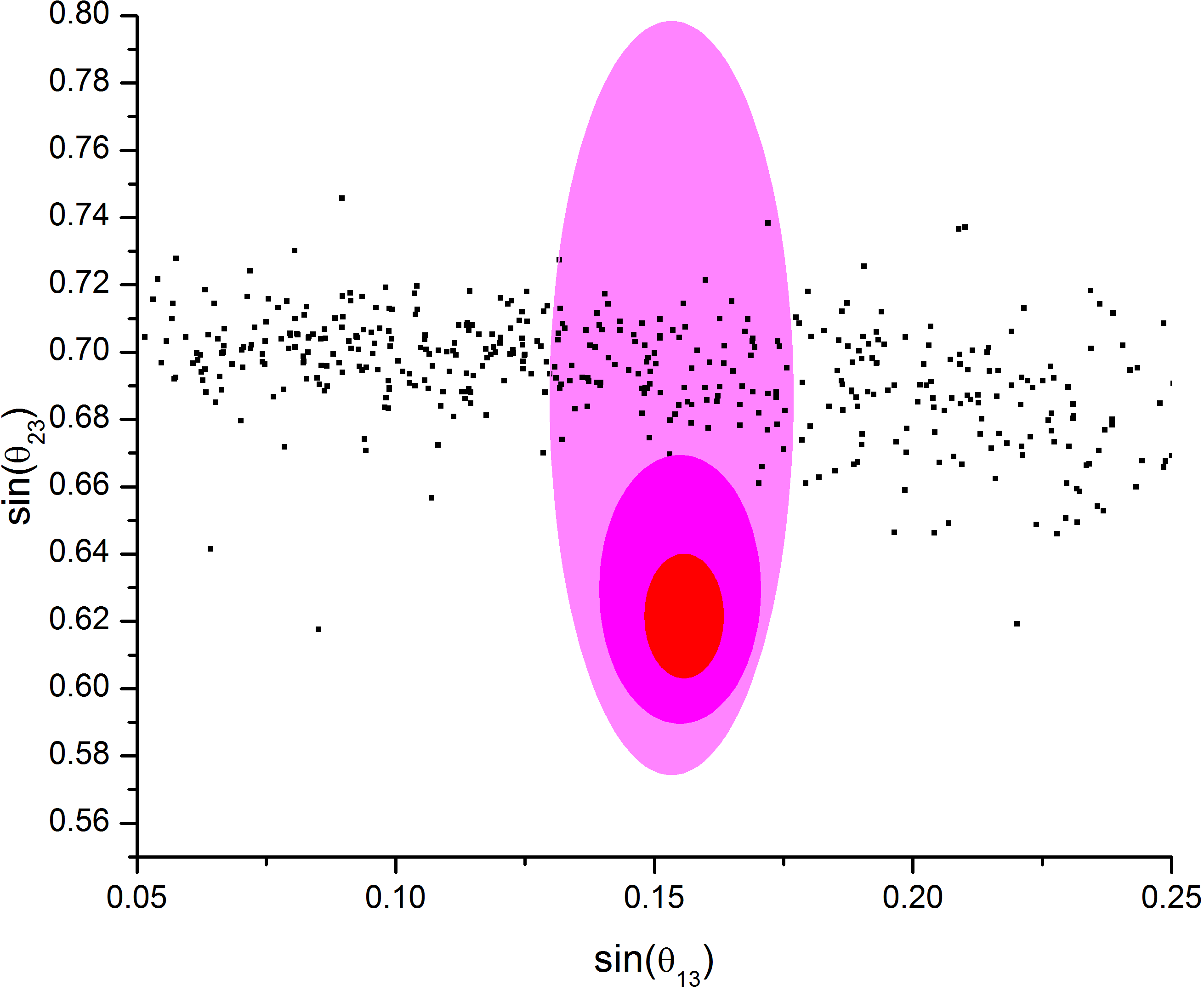}
\caption{\label{fig:expt-compat} Graphs of $\sin(\theta_{12})$ and $\sin(\theta_{23})$ against $\sin(\theta_{13})$. The error ellipses for the mixing angles are roughly derived from the $1\sigma$ to $3\sigma$ ranges in \cite{ref:expt}.}
\end{figure}

Figure~\ref{fig:expt-compat} shows plots of $\sin(\theta_{12})$ and $\sin(\theta_{23})$ against $\sin(\theta_{13})$, using parameter sets from $\mathcal{C}_2$ and zoomed into the regions around $\sin(\theta_{13}) \sim 0.15$. We see that for a large $\sin(\theta_{13}) \sim 0.15$, $\sin(\theta_{12})$ is fairly evenly distributed between $0.45$ and $0.7$, with about $25\%$ of the points within the $3\sigma$ range, as opposed to a bimodal distribution peaked at the two extremes. We hence conclude that corrections to $\theta_{12}$ need not be of the same size as $\theta_{13}$, and so this model can be made compatible with all three experimental mixing angles.

%
%
\section{Modifying the flavon vacuum alignment} \label{sect:second-method}

In this section, we consider the idea of changing the alignments of the flavon VEVs to obtain a nonzero $\theta_{13}$. In particular, we focus on flavons associated with the charged lepton masses, and show that the effects on $\theta_{13}$ are enhanced by a factor that scales with $m_\tau/m_\mu$. We assume that the corrections discussed in the previous section are not important, e.g. when $\epsilon$ is extremely small\footnote{Actually even for very small $\epsilon$, the zeroth-order approximation is really $M_l = \frac{v_H}{\Lambda} \left( A - \frac{y^T_m}{y^{T\prime}_m} C \right)$. However, since $A$ and $C$ are of the same form, this is equivalent to a different choice of Yukawas for $A$ in $M_l = \frac{v_H}{\Lambda} A$. Henceforth, for notational simplicity, we ignore the $C$ correction.}, so that $M_l \propto A$. With a modified alignment, $A$ is no longer given by Eq.~(\ref{eqn:form-of-a}). In particular, it is not of the form $M_{\text{aligned}}$ and hence a nonzero $\theta_{13}$ can be generated. We do not attempt to explain the origin of the modified alignment, and will just focus on the consequence of such a modification.

In general, the relative alignments between all the flavons can be varied. However, we can illustrate most of the important features by just varying the alignment of $\langle \phi \rangle$:
\begin{equation}
\langle \phi \rangle = \sqrt{3} v \left( \begin{array}{c} \sin(a)\cos(b) \\ \sin(a)\sin(b) \\ \cos(a) \end{array} \right).
\end{equation}
We recover the original alignment when $a = \arcsin\left(\sqrt{2/3}\right)$ and $b = \frac{\pi}{4}$. With this alignment we now have
\begin{equation}
A = \sqrt{3} \left( \begin{array}{ccc}
 y_e v s_a c_b &  y_m v s_a c_b + \frac{y^5_m v_5}{\sqrt{3}}(\omega^2-\omega) &  y_m v s_a c_b + \frac{y^5_m v_5}{\sqrt{3}}(\omega-\omega^2)\\
 y_e v s_as_b &  y_m v s_as_b \omega + \frac{y^5_m v_5}{\sqrt{3}}(1-\omega^2) &  y_m v s_as_b \omega^2 + \frac{y^5_m v_5}{\sqrt{3}}(1-\omega)\\
y_e v c_a &  y_m v c_a \omega^2 + \frac{y^5_m v_5}{\sqrt{3}}(\omega-1) &  y_m v c_a \omega + \frac{y^5_m v_5}{\sqrt{3}}(\omega^2-1)\,,
\end{array} \right) \label{eqn:new-alignment}
\end{equation}
where $s_x\equiv \sin x$ and $c_x \equiv \cos x$. The angle between the original and modified alignment, which we denote as $\chi$, can be thought of as the small parameter in this approach. At first glance, we might expect the size of $\theta_{13}$ to be given by $\chi$. However, the near-degeneracy of $m_e$ and $m_\mu$ relative to $m_\tau$ again comes into play, and so $\theta_{13}$ is amplified by a factor of $O(m_\tau/m_\mu)$. As discussed in Sec.~\ref{sect:amplification}, the amplification is not $O(m_\tau^2/m_\mu^2)$ since $M_l (M_l)^\dagger$ is obviously factorizable at a well-defined order in $\chi$.

As an aside, it is not particularly difficult to perform a parameter scan to find values of VEVs, Yukawas and alignment angles that generate the correct mass spectrum and mixing angles. A useful observation is that the relation
\begin{equation}
m^2_e+m^2_\nu+m^2_\tau = {\rm Tr}\left[M_l (M_l)^\dagger\right] = 3\frac{v^2_Hv^2}{\Lambda^2}(|y_e|^2+2|y_m|^2)+18\frac{v^2_H v^2_5}{\Lambda^2}|y^5_m|^2,
\end{equation}
is independent of the alignment, hence allowing us to reduce the number of parameters by one. However, such a scan is not very useful, since the effects of block diagonalization discussed in the previous section are expected to be significant given the constraints on the hierarchy of energy scales. Therefore, we will only focus on demonstrating the enhancement effects.

We compute the tree-level \UPMNS for four large collections of random parameter sets $\mathcal{C}_3$, $\mathcal{C}_4$, $\mathcal{C}_5$ and $\mathcal{C}_6$. Details of their generation are provided in Appendix~\ref{sect:random2}. The collections differ in the conditions imposed on the ratio of mass eigenvalues: No conditions have been imposed on $\mathcal{C}_3$, so the mass eigenvalues are typically of the same order, while the correct mass ratios have been imposed on $\mathcal{C}_4$. The conditions on mass ratios $m_\mu^2/m_\tau^2$ and $m_e^2/m_\tau^2$ have been (unphysically) modified to be $10$ times smaller in $\mathcal{C}_5$, and $100$ times smaller in $\mathcal{C}_6$. In other words, the mass ratio $m_\tau/m_\mu$ relevant to the enhancement effect is larger in $\mathcal{C}_5$ and even more so in $\mathcal{C}_6$.

\begin{figure}[t]
\centering
\includegraphics[width=.49\linewidth]{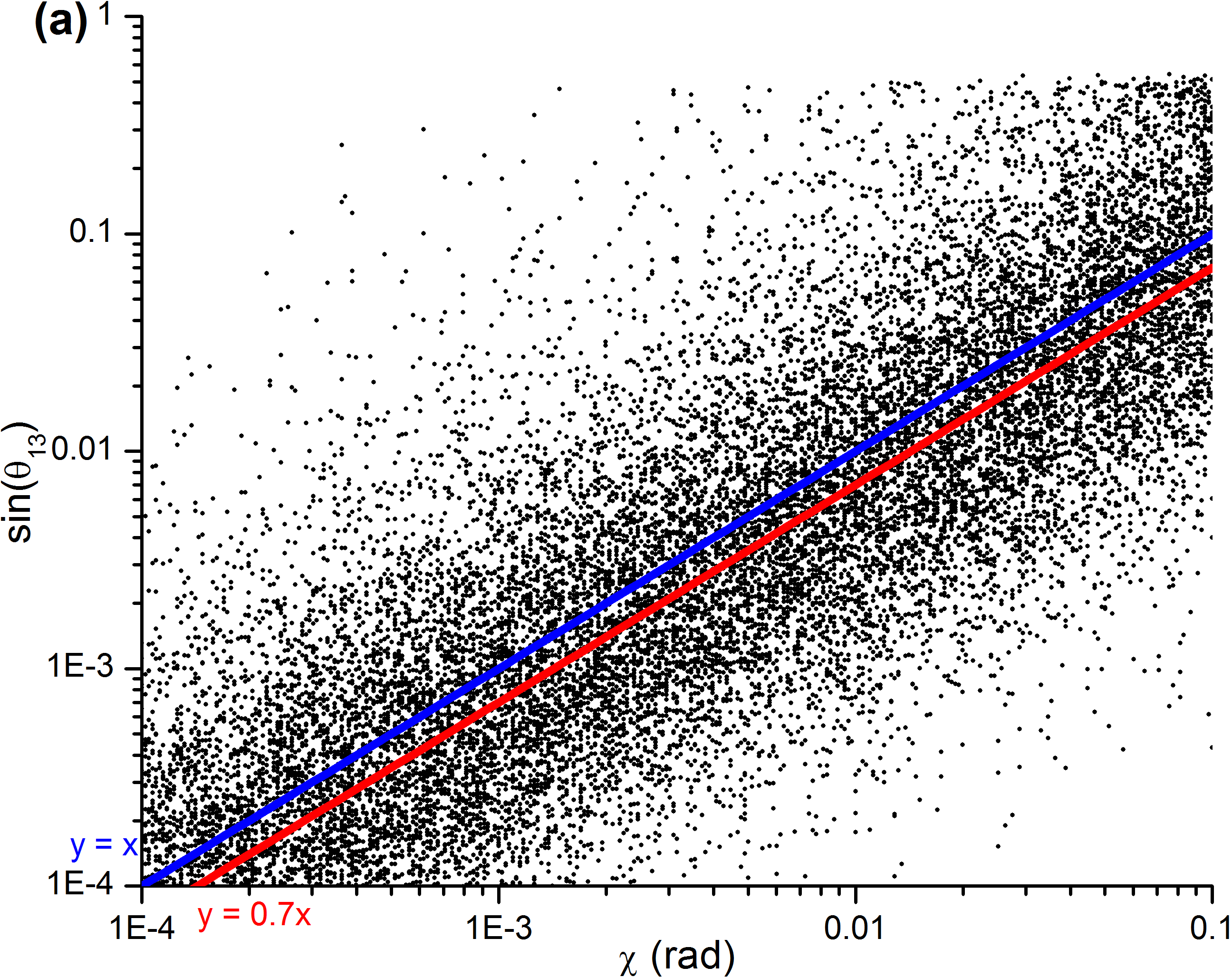}
\includegraphics[width=.49\linewidth]{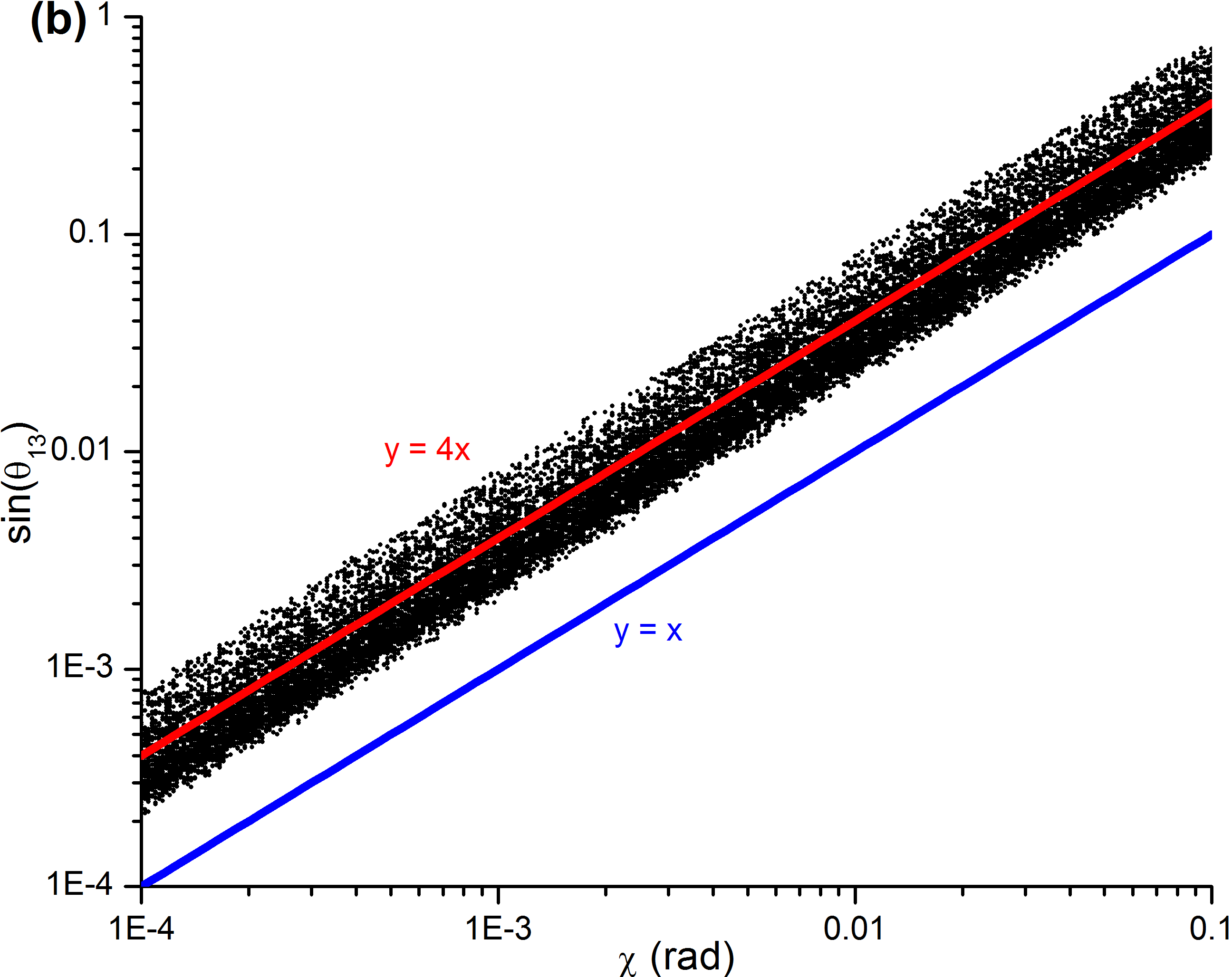}
\includegraphics[width=.49\linewidth]{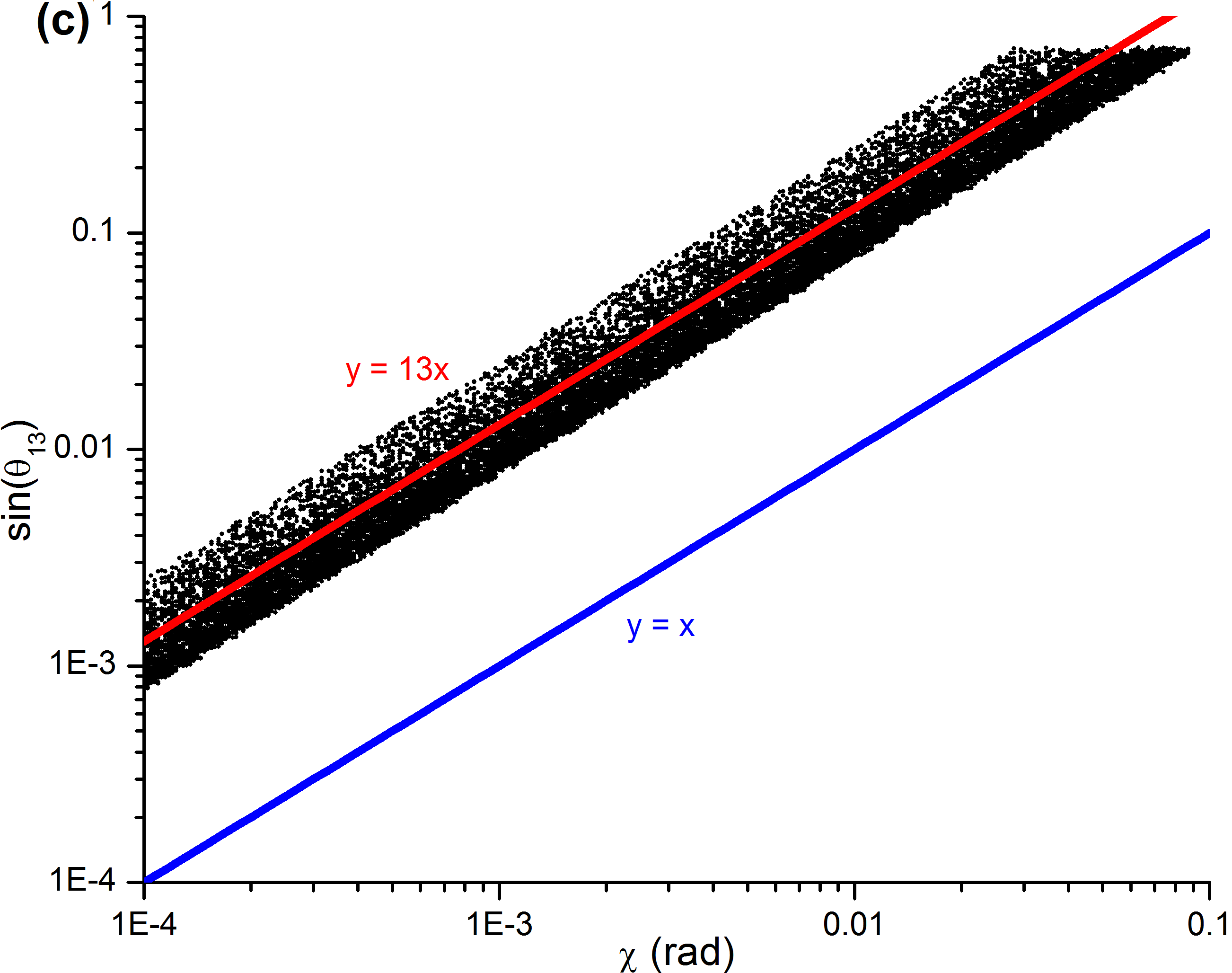}
\includegraphics[width=.49\linewidth]{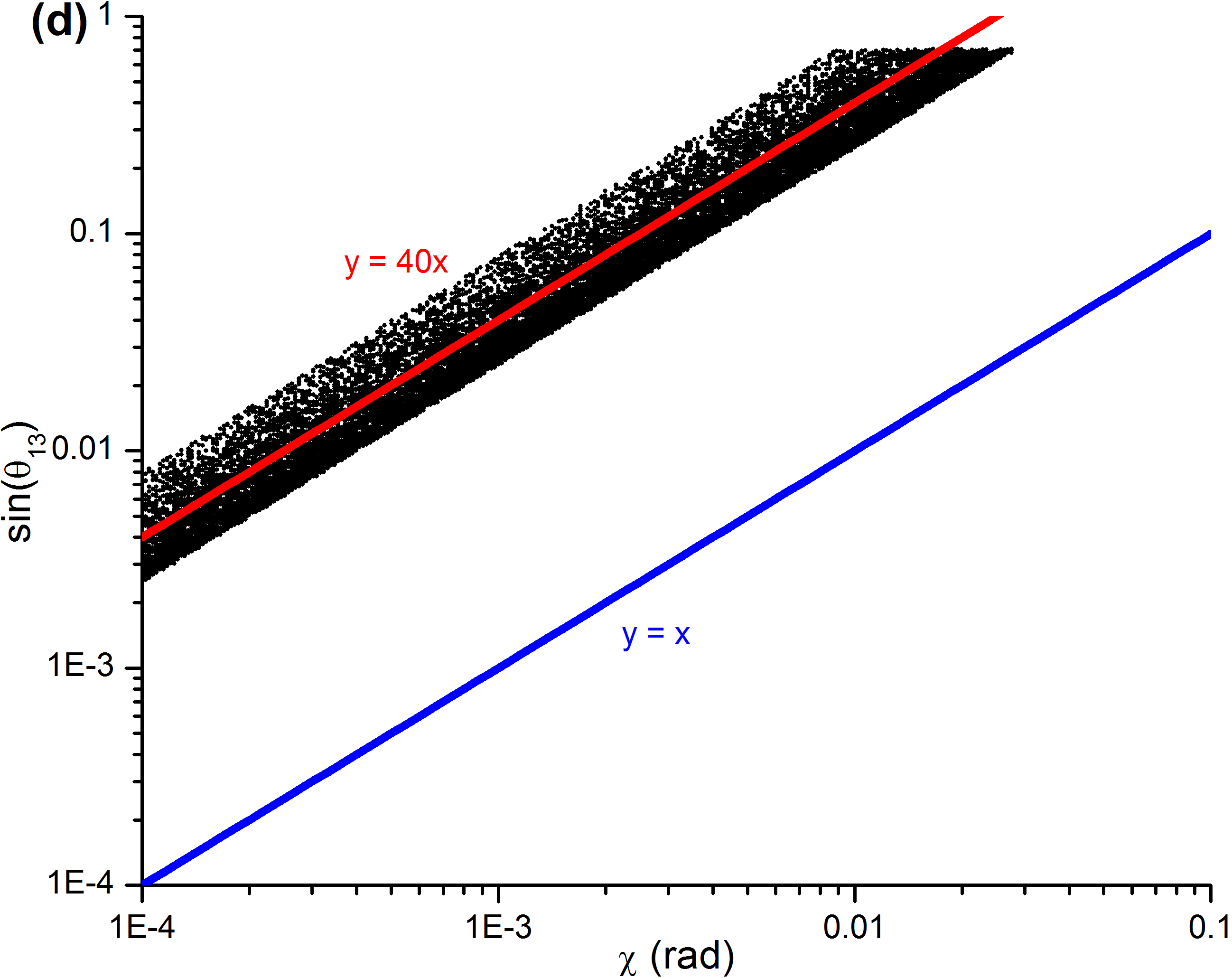}
\caption{\label{fig:second-method} Graphs of $\sin(\theta_{13})$ against the change in flavon alignment $\chi$ for collections (a) $\mathcal{C}_3$, (b) $\mathcal{C}_4$, (c) $\mathcal{C}_5$ and (d) $\mathcal{C}_6$. The collections differ in the conditions imposed on the ratio of mass eigenvalues.}
\end{figure}

Figure~\ref{fig:second-method} shows the graphs of $\sin(\theta_{13})$ against $\chi$ for all the four collections. Just as in Section~\ref{sect:simulation1}, we observe an enhancement effect in $\mathcal{C}_4$ relative in $\mathcal{C}_3$, although it is smaller than the predicted enhancement of $m_\tau/m_\mu$ due to large $O(1)$ factors that we have not taken into account. Nonetheless, the graphs for $\mathcal{C}_5$ and $\mathcal{C}_6$ clearly demonstrate that the enhancement factor scales as $m_\tau/m_\mu$, in agreement with our predictions.

To conclude, we have demonstrated that modifying the alignment of flavons associated with charged lepton masses give rise to a nonzero $\theta_{13}$, with an enhancement factor that scales as $m_\tau/m_\mu$. This enhancement may be applicable to a large class of $A_4$ models since the only feature we have alluded to beyond the minimal $A_4$ model are the additional mass terms involving $v_5$, which can be easily reproduced with an additional $A_4$ flavon.

%
%

\section{Discussion and conclusion.} \label{sect:conclusion}

Having discussed the two approaches of obtaining a nonzero $\theta_{13}$, there remains various issues that we did not touch on and are worth further investigation. First, we have not addressed one shortcoming of the model originally mentioned in \cite{ref:berger}. This is the issue of Goldstone bosons when the global $SO(3)_F$ symmetry is broken, and the issue of mixed anomalies involving $U(1)_Y$ should we gauge the $SO(3)_F$ symmetry to eat these Goldstone bosons. However, the variety of modified models in Appendix~\ref{app:z-n-symm} suggests that it should be possible to introduce additional heavy leptons to address the issue of anomalies, and yet suppress their mass couplings to the existing leptons using auxiliary symmetries.

Second, our analysis so far is only at the classical level. We have yet to consider the running of parameters down to the electroweak scale \cite{ref:radiative1, ref:radiative2, ref:radiative3, ref:radiative4, ref:radiative5, ref:radiative6, ref:radiative7, ref:radiative8, ref:radiative9, ref:radiative10, ref:radiative11, ref:radiative12, ref:radiative13}. Nonetheless, since our neutrino mass spectrum is not quasi-degenerate, the classical results should hold as a first approximation.

Last is the issue of fine-tuning of the charged-lepton masses. In the minimal $A_4$ model, this can be resolved by modifying the model to give naturally small electron Yukawas. In the $SO(3)_F \rightarrow A_4$ model however, suppressing particular Yukawas do not guarantee the correct mass hierarchy, since the subleading corrections from block diagonalization can significantly affect the small eigenvalues. Still, we have demonstrated with our simulation in Section~\ref{sect:simulation1} that small electron masses can be achieved, although the small fraction of successful parameter sets imply that specific relations between the Yukawas are required. Unfortunately, the exact forms of these relations are far from obvious, hence obscuring any UV explanation of the fine-tuning.

To conclude, the $SO(3)_F \rightarrow A_4$ model of \cite{ref:berger} is the UV completion of an effective $A_4$ model with the purpose of reproducing the tri-bimaximal mixing pattern in \UPMNS. However, due to mixing between heavy and SM charged leptons, we find that the model actually predicts a nonzero $\theta_{13}$, with the size of $\theta_{13}$ a measure of the ratio of the $A_4$-breaking to $SO(3)_F$-breaking scales. We have also shown that this model can reproduce both the measured light lepton spectrum and the mixing angles, and is hence compatible with experimental observations. Nonetheless, there exists various unattractive aspects of the model, in particular the fine-tuning of the charged-lepton eigenvalues and the need of an auxiliary symmetry on top of the original $SO(3)_F$ symmetry. We hope to resolve these issues in a future work.

%
%

\section*{Acknowledgments}
We thank Josh Berger and Jeff Dror for helpful discussions. 
YG is a Weston Visiting Professor at the Weizmann Institute.
This work was partially supported by a grant from the Simons Foundation ($\#$267432 to Yuval Grossman).
The work of YG is supported is part by the U.S. National Science Foundation through grant PHY-0757868 and by the United States-Israel Binational Science Foundation (BSF) under grant No.~2010221.

%
%

\appendix

\section{Modified models with similar SM lepton phenomenology} \label{app:z-n-symm}

\subsection{Overview}

As pointed out in Section~\ref{sect:lagrangian}, there are two issues with the Lagrangian given by Eq.~(\ref{eqn:charged-lepton-lagrangian}) and (\ref{eqn:neutrino-lagrangian}). First, it is not the most general one consistent with $SU(2)_L \times U(1)_Y$ gauge and global $SO(3)_F$ symmetries. Second, it is not clear whether the truncation of the Lagrangian is consistent with our hierarchy of scales. As an example, we have omitted dimension-six terms like $-\overline{\psi_l}^a \psi_m^{bc} \epsilon^{bde} H T^{adf} T^{cef} / \Lambda^2$ while keeping dimension-five terms like $-\overline{\psi_l}^a \psi_m^{ab} H \phi^b/\Lambda$, both of which contribute to the Dirac mass of $\psi_l$ by an amount $\sim v_H v_T^2/\Lambda^2$ and $\sim v_H v/\Lambda$ respectively. However, since $\{v, v_5, v'\} \ll v_T \ll \Lambda$, it is not immediately clear that the former is necessarily smaller than the latter, unless we impose the additional restriction that $v_T/\Lambda \ll \{v, v_5, v'\}/v_T$.

It turns out that both issues can be addressed if we modify the model to include an auxiliary $Z_n$ symmetry and a $Z_n$ flavon $S$. The modified model is designed to reproduce the same lepton mass matrices as the original Lagrangian. The auxiliary $Z_n$ symmetry forbids lower-dimension terms otherwise allowed by the gauge and $SO(3)_F$ symmetry that would have changed the mass matrices, as well as certain higher-dimension terms (such as those quadratic in $T$) that if neglected, would have led to large truncation errors. The flavon $S$ is a gauge and $SO(3)_F$ singlet, the VEV of which is related to the neutrino Majorana mass parameter $M$.

Two versions of modified models are discussed below, the main difference being the effective size of $y_\nu$ in the original Lagrangian, and hence the neutrino seesaw scale.

\subsection{Model 1: $Z_8$, with typical seesaw scale}

We assign the following $Z_8$ representations to the matter fields:
\begin{table}[H]
\renewcommand{\tabcolsep}{8pt}
\centering
\begin{tabular} {| c | c | c | c | c | c | c | c | c | c | c | c |}
\hline
Field & $\psi_l$ & $\psi_f$ & $\psi_e$ & $\psi_m$ & $\psi_n$ & $H$ & $\phi$ & $\phi'$ & $\phi_5$ & $S$ & $T$\\
\hline
$Z_8$ rep. & $e^{i \frac{\pi}{4}}$ & $e^{i \frac{\pi}{4}}$ & $e^{-i \frac{3\pi}{4}}$ & $e^{-i \frac{3\pi}{4}}$ & $e^{i \frac{\pi}{4}}$ & $+1$ & $-1$ & $+i$ & $-1$ & $-i$ & $-1$\\
\hline
\end{tabular}
\end{table}
Note that $\phi'$ and $S$ have to be complex fields since they are in complex $Z_8$ representations.

For the charged-lepton sector, since Dirac masses for $\psi_f^-$ and $\psi_l^-$ are generated by operators that are at least dimension-four and five respectively, with the latter always requiring a Higgs field, it is natural to use these minimum criteria as the truncation scheme. The most general Lagrangian turns out to be same as the original $\mathcal{L}_e$ given in Eq.~(\ref{eqn:charged-lepton-lagrangian}). The higher-dimension terms we have neglected are given heuristically (coefficients and $SO(3)$ indices suppressed) by
\begin{equation}
\begin{aligned}
\mathcal{L}^{\text{h.o.}}_e
\sim
&- \frac{1}{\Lambda} \left( \overline{\psi_f}\psi_e \phi'S^* + \overline{\psi_f}\psi_m \phi' \phi' + ... \right)\\
&- \frac{1}{\Lambda^2} \left( \overline{\psi_l}\psi_e H \phi' S^* + \overline{\psi_l}\psi_m H \phi' \phi' + ... + \overline{\psi_f}\psi_e TTT + \overline{\psi_f}\psi_m TTT + ... \right)\\
&- \frac{1}{\Lambda^3} \left( \overline{\psi_l}\psi_e HTTT + \overline{\psi_l}\psi_m HTTT + ... \right)- ... + \text{h.c.}
\end{aligned}
\end{equation}
We note that terms like $-\overline{\psi_f}\psi_m TT/\Lambda$ that may lead to large truncation errors (if neglected) are explicitly forbidden by the $Z_8$ symmetry.
 
We now discuss the neutrino sector. Neutrino Dirac masses are generated by operators that are at least dimension four and always require a Higgs field. While neutrino Majorana masses can be generated by dimension-four operators, we also allow dimension-five operators that can potentially give comparable contributions as a result of the hierarchy of scales. With the above as the truncation scheme, the neutrino Lagrangian is then given by $\mathcal{L}^{\text{l.o.}}_\nu = \mathcal{L}^{\text{l.o.(old)}}_\nu + \mathcal{L}^{\text{l.o.(new)}}_\nu$, where
\begin{equation}
\begin{aligned}
\mathcal{L}^{\text{l.o.(old)}}_\nu &= -x^S_\nu S \overline{ \psi^{\text{c}}_n}^a \psi^a_n - x_\nu \frac{1}{\Lambda} \overline{\psi^{\text{c}}_n}^a \psi^b_n \phi^{\prime c} T^{abc} - y_\nu \overline{\psi_l}^a \tilde{H} \psi^a_n + \text{h.c.} \\
\mathcal{L}^{\text{l.o.(new)}}_\nu &\sim -\frac{1}{\Lambda} \left( \overline{\psi^{\text{c}}_n} \psi_n \phi' \phi + \overline{\psi^{\text{c}}_n} \psi_n \phi' \phi_5 \right) + \text{h.c.}
\end{aligned}
\end{equation}
The higher dimension terms that we have neglected are
\begin{equation}
\mathcal{L}^{\text{h.o.}}_\nu
\sim
- \frac{1}{\Lambda^2} \left( \overline{\psi_l}\psi_n HTT + ... + \overline{\psi^{\text{c}}_n} \psi_n STT + ... \right) - ... + \text{h.c.}
\end{equation}

When the flavon $S$ gains a VEV $v_S$, if we identify $x^S_\nu v_S$ with $M$, $\mathcal{L}^{\text{l.o.(old)}}_\nu$ then generates the same neutrino mass matrix as the original Lagrangian. Therefore, the largest contributions that we have omitted from our original mass matrix come from $\mathcal{L}^{\text{l.o.(new)}}_\nu$ and $\mathcal{L}^{\text{h.o.}}_\nu$. Note that $\mathcal{L}^{\text{l.o.(old)}}_\nu$ cannot be eliminated through a different implementation of auxiliary symmetries without significantly modifying the charged-lepton mass matrix.

We now analyze the fractional errors in both the charged lepton and neutrino mass matrices as a result of the various omitted contributions discussed above. For simplicity, we assume that all the Yukawas of terms that contribute to the same mass type, omitted or otherwise, are of the same order. As a result, the Yukawas (heuristically denoted as $y$) cancel out in the fractional errors, which we summarize in the table below. Note that we have defined $\epsilon_T \equiv v_T/\Lambda$.

\begin{table}[H]
\renewcommand{\tabcolsep}{8pt}
\centering
\begin{tabular} {| c | c | c | c |}
\hline
Mass types & Smallest contributions & Largest contributions & Fractional\\
& included & omitted & error\\
\hline
$\psi_l^-$, Dirac & $y\frac{v_H v}{\Lambda}$ & $\sup\lbrace y\frac{v_H v^2}{\Lambda^2}, y\frac{v_H v_T^3}{\Lambda^3}\rbrace$ & $\sup\lbrace \epsilon \epsilon_T, \frac{\epsilon_T^2}{\epsilon} \rbrace$\\
$\psi_f^-$, Dirac & $yv$ & $\sup\lbrace y\frac{v^2}{\Lambda}, y\frac{v_T^3}{\Lambda^2}\rbrace$ & $\sup\lbrace\epsilon \epsilon_T, \frac{\epsilon_T^2}{\epsilon}\rbrace$\\
Neutrino, Dirac & $yv_H$ & $y\frac{v_H v_T^2}{\Lambda^2}$ & $\epsilon_T^2$\\
Neutrino, Majorana & $y\frac{vv_T}{\Lambda}$ & $y\frac{v^2}{\Lambda}$ & $\epsilon$ \\
\hline
\end{tabular}
\end{table}
We want all fractional errors to be smaller than $\epsilon$ so that the omitted contributions generate smaller corrections to $\theta_{13}$ than what we have discussed in Section~\ref{sect:first-method}. Except for neutrino Majorana masses, this can be achieved by choosing a hierarchy where $\epsilon_T < \epsilon$. For neutrino Majorana masses, fine-tuning may be required to suppress the Yukawas associated with the omitted contributions to reduce the fractional errors. We have not taken into account any enhancement effects which may ameliorate or exacerbate the fine-tuning.

\subsection{Model 2: $Z_8$, with lower seesaw scale}

In this model, we assign different $Z_8$ representations to the matter fields.
\begin{table}[H]
\renewcommand{\tabcolsep}{8pt}
\centering
\begin{tabular} {| c | c | c | c | c | c | c | c | c | c | c | c |}
\hline
Field & $\psi_l$ & $\psi_f$ & $\psi_e$ & $\psi_m$ & $\psi_n$ & $H$ & $\phi$ & $\phi'$ & $\phi_5$ & $S$ & $T$\\
\hline
$Z_8$ rep. & $-1$ & $-1$ & $+1$ & $+1$ & $e^{i \frac{\pi}{4}}$ & $+1$ & $-1$ & $+i$ & $-1$ & $e^{i \frac{3\pi}{4}}$ & $-1$\\
\hline
\end{tabular}
\end{table}
Again, we note that $\phi'$ and $S$ have to be complex fields. The charged-lepton Lagrangian is the same as the one in the previous model. For the neutrino sector, since neutrino Dirac masses now only arise at dimension five, the truncation scheme is modified accordingly. The neutrino Lagrangian is given by $\mathcal{L}^{\text{l.o.}}_\nu = \mathcal{L}^{\text{l.o.(old)}}_\nu + \mathcal{L}^{\text{l.o.(new)}}_\nu$, where
\begin{equation}
\begin{aligned}
\mathcal{L}^{\text{l.o.(old)}}_\nu &= -x^S_\nu \frac{1}{\Lambda} S^2 \overline{ \psi^{\text{c}}_n}^a \psi^a_n - x_\nu \frac{1}{\Lambda} \overline{\psi^{\text{c}}_n}^a \psi^b_n \phi^{\prime c} T^{abc} - y'_\nu \frac{1}{\Lambda} S \overline{\psi_l}^a \tilde{H} \psi^a_n\\
\mathcal{L}^{\text{l.o.(new)}}_\nu &\sim -\frac{1}{\Lambda} \left( \overline{\psi^{\text{c}}_n} \psi_n \phi' \phi + \overline{\psi^{\text{c}}_n} \psi_n \phi' \phi_5 \right) + \text{h.c.}
\end{aligned}
\end{equation}
The higher dimension terms that we have neglected are
\begin{equation}
\mathcal{L}^{\text{h.o.}}_\nu
\sim
- \frac{1}{\Lambda^2} \left( \overline{\psi^{\text{c}}_n} \psi_n S^* S^* T + \overline{\psi^{\text{c}}_n} \psi_n \phi^{\prime*} T T + ... \right) 
- \frac{1}{\Lambda^3} \left( \overline{\psi_l}\psi_n HSTT + ... \right)
- ... + \text{h.c.}
\end{equation}

When the flavon $S$ gains a VEV $v_S$, if we identify $x^S_\nu v_S^2 / \Lambda$ with $M$ and $y'_\nu v_S /\Lambda$ with $y_\nu$, $\mathcal{L}^{\text{l.o.(old)}}_\nu$ then generates the same neutrino mass matrix as the original Lagrangian, but with $y_\nu$ naturally suppressed by $v_S/\Lambda$. This allows for a seesaw scale $M$ that is roughly two orders of magnitude lower than usual.

The fractional errors for the different mass types are given in the table below. Again the preferred hierarchy is one where $\epsilon_T < \epsilon$, and fine-tuning is still required to suppress the Yukawas associated with neutrino Majorana mass contributions that have been omitted.
\begin{table}[H]
\renewcommand{\tabcolsep}{8pt}
\centering
\begin{tabular} {| c | c | c | c |}
\hline
Mass types & Smallest contributions & Largest contributions & Fractional\\
 & included & omitted & error \\
\hline
$\psi_l^-$, Dirac & $\frac{v_H v}{\Lambda}$ & $\sup \lbrace \frac{v_H v^2}{\Lambda^2}, \frac{v_H v_T^3}{\Lambda^3} \rbrace$ & $\sup \lbrace \epsilon \epsilon_T, \frac{\epsilon_T^2}{\epsilon} \rbrace$\\
$\psi_f^-$, Dirac & $v$ & $\sup \lbrace \frac{v^2}{\Lambda}, \frac{v_T^3}{\Lambda^2} \rbrace$ & $\sup \lbrace \epsilon \epsilon_T, \frac{\epsilon_T^2}{\epsilon} \rbrace$\\
Neutrino, Dirac & $\frac{v_H \sqrt{v v_T}}{\Lambda}$ & $\frac{v_H \sqrt{v v_T} v_T^2}{\Lambda^3}$ & $\epsilon_T^2$\\
Neutrino, Majorana & $\frac{vv_T}{\Lambda}$ & $\sup \lbrace \frac{v^2}{\Lambda}, \frac{v v_T^2}{\Lambda^2} \rbrace$ & $\sup \lbrace \epsilon, \epsilon_T \rbrace$ \\
\hline
\end{tabular}
\end{table}

%
%

\section{Nonunitary factors in \UPMNS} \label{app:nonunitary}

In this Appendix, we discuss the origin of nonunitary factors mentioned in Section~\ref{sect:review} and why they turn out to be negligible. The charged current weak interaction acts between the left-handed SM charged leptons and neutrinos, both of which are linear combinations of light and heavy mass eigenstates. \UPMNS characterizes this interaction between only the light mass eigenstates.

Define $6 \times 6$ unitary matrices $U_{l,\text{full}}^{6 \times 6}$ and $U_{l,\text{full}}^{6 \times 6}$ that are required to fully diagonalize $M^{6 \times 6}_l (M^{6 \times 6}_l)^\dagger$ and $M^{6 \times 6}_\nu$:
\begin{align}
U_{\nu, \text{full}}^{6 \times 6} M^{6 \times 6}_\nu (U_{\nu, \text{full}}^{6 \times 6})^T &= \left( \begin{array}{cccccc}
m_1 & 0 & 0 & 0 & 0 & 0\\
0 & m_2 & 0 & 0 & 0 & 0\\
0 & 0 & m_3 & 0 & 0 & 0\\
0 & 0 & 0 & m_{1'} & 0 & 0\\
0 & 0 & 0 & 0 & m_{2'} & 0\\
0 & 0 & 0 & 0 & 0 & m_{3'}\\
\end{array} \right),\\
U_{l, \text{full}}^{6 \times 6} M^{6 \times 6}_l (M^{6 \times 6}_l )^\dagger (U_{l, \text{full}}^{6 \times 6})^\dagger &= \left( \begin{array}{cccccc}
m_e^2 & 0 & 0 & 0 & 0 & 0\\
0 & m_\mu^2 & 0 & 0 & 0 & 0\\
0 & 0 & m_\tau^2 & 0 & 0 & 0\\
0 & 0 & 0 & m_{e'}^2 & 0 & 0\\
0 & 0 & 0 & 0 & m_{\mu'}^2 & 0\\
0 & 0 & 0 & 0 & 0 & m_{\tau'}^2\\
\end{array} \right),
\end{align}
where $'$ indicates a heavy lepton. We can write $U_{l,\text{full}}^{6 \times 6}$ and $U_{l,\text{full}}^{6 \times 6}$ in terms of $3 \times 3$ blocks as shown here:
\begin{equation}
U_{l,\text{full}}^{6 \times 6} = \left( \begin{array}{cc} U_{l,\text{full}} & U'_{l,\text{full}} \\ U''_{l,\text{full}} & U'''_{l,\text{full}} \end{array}\right), \quad
U_{\nu,\text{full}}^{6 \times 6} = \left( \begin{array}{cc} U_{\nu,\text{full}} & U'_{\nu,\text{full}} \\ U''_{\nu,\text{full}} & U'''_{\nu,\text{full}} \end{array}\right).
\end{equation}
\UPMNS is then given by
\begin{equation}
U_{\text{PMNS}} = U_{l,\text{full}}(U_{\nu, \text{full}})^\dagger,
\end{equation}
Since the $3 \times 3$ blocks are nonunitary in general, we expect the same for \UPMNS.

It is perhaps more illustrative to regard the diagonalization as a two-step process, which we demonstrate here with the neutrino sector. Define a $6 \times 6$ unitary matrix $U_{\nu,\text{bd}}^{6 \times 6}$ that is required to block-diagonalize $M^{6 \times 6}_\nu$:
\begin{equation}
U_{\nu,\text{bd}}^{6 \times 6} M^{6 \times 6}_\nu (U_{\nu,\text{bd}}^{6 \times 6})^T = \left( \begin{array}{cc} M_\nu&0\\0&M_{\nu'}\end{array} \right),
\end{equation}
where $M_\nu$ and $M_{\nu'}$ are the $3 \times 3$ Majorana mass matrices for the light and heavy neutrinos. Again we can write $U_{\nu,\text{bd}}^{6 \times 6}$ in terms of $3 \times 3$ blocks:
\begin{equation}
U_{\nu,\text{bd}}^{6 \times 6} = \left( \begin{array}{cc} U_{\nu,\text{bd}} & U'_{\nu,\text{bd}} \\ U''_{\nu,\text{bd}} & U'''_{\nu,\text{bd}} \end{array}\right).
\end{equation}
Let $U_\nu$ be the $3 \times 3$ unitary matrix required to diagonalize $M_\nu$:
\begin{equation}
U_\nu M_\nu (U_\nu)^T = \left( \begin{array}{ccc} m_1 & 0 & 0 \\ 0 & m_2 & 0 \\ 0 & 0 & m_3 \end{array} \right),
\end{equation}
We can then show that
\begin{equation} U_{\nu,\text{full}} = U_\nu U_{\nu,\text{bd}}. \end{equation}
In other words, $U_{\nu,\text{full}}$ can be decomposed into a unitary factor associated with the diagonalization of $M_\nu$, and a non-unitary factor associated with the block-diagonalization of $M^{6 \times 6}_\nu$. Applying a similar two-step process to the charged lepton sector gives us the factorization
\begin{equation} U_{l,\text{full}} = U_l U_{l,\text{bd}}. \end{equation}
\UPMNS is then given by
\begin{equation}
U_{\text{PMNS}} = U_l U_{l,\text{bd}} (U_{\nu,\text{bd}})^\dagger (U_\nu)^\dagger.
\end{equation}
This expression differs from Eq.~(\ref{eqn:naive-PMNS}) by the nonunitary factor $U_{l,\text{bd}} (U_{\nu,\text{bd}})^\dagger$ associated with the block-diagonalization process. However, we can show that $U_{l,\text{bd}}$ and $(U_{\nu,\text{bd}})^\dagger$ deviate from the identity matrix by terms of order $O(\frac{v^2_H}{\Lambda^2})$ and $O(\frac{v^2_H}{M^2})$ respectively. Based on the energy scales in Eq.~(\ref{eqn:scales}), these are exceedingly small deviations. Hence, their effects on \UPMNS are negligible and \UPMNS can be considered to be unitary. 

%
%

\section{Generation of random parameter sets}

In this appendix, we discuss how we generate the various collections of random parameter sets used in the simulations.

\subsection{$\mathcal{C}_1$ and $\mathcal{C}_2$} \label{app:random}

The collections $\mathcal{C}_1$ and $\mathcal{C}_2$ are used in Figure~\ref{fig:first-method}. In each collection, the VEV $v_T$ is a log flat random variable between $10^{16}$ and $10^{19} \, \text{GeV}$, while $v$ and $v_5$ are uniform random variables between $10^{15}$ and $10^{16} \, \text{GeV}$.

In $\mathcal{C}_1$, which consists of $20,000$ sets, all eight charged-lepton Yukawas are simply $O(1)$ uniform random complex variables, with real and imaginary parts between $-3$ and $3$. In $\mathcal{C}_2$, we want to restrict the parameter sets to only those that produce the correct charged-lepton mass ratios. Ideally, we would like to use the same definitions of random variables as $\mathcal{C}_1$, and simply reject the parameter sets that fail the cut. However, the very small measure of the allowed parameter space makes this computationally prohibitive, so we instead adopt an alternative procedure for $\mathcal{C}_2$ which we outline below.

First, we define two new uniform random complex variables $\alpha_1$ and $\alpha_2$ of magnitudes $O(\frac{1}{1000})$ and $O(\frac{1}{10})$ respectively, that satisfy the relations
\begin{align}
y'_e =& \frac{y^{T\prime}_m}{y^T_m} \left( y_e - \alpha_1 \right), \label{eqn:yukawa-constraints1} \\
y^{5\prime}_m =& \frac{y^{T\prime}_m}{y^T_m} y^5_m + \frac{y^{T\prime}_m}{y^T_m} \frac{i}{\sqrt{3}} \frac{v}{v_5} \left(y_m - \frac{y^T_m}{y^{T\prime}_m} y'_m \right) \left( 1 - \alpha_2 \right) \label{eqn:yukawa-constraints2}
\end{align}
Instead of generating all eight charged-lepton Yukawas randomly, we now generate only six of them (excluding $y'_e$ and $y^{5\prime}_m$), together with $\alpha_1$ and $\alpha_2$. $y'_e$ and $y^{5\prime}_m$ are then obtained from the relations above. Since we still want all Yukawas to be $O(1)$, we discard the parameter set should the resulting $y'_e$ and $y^{5\prime}_m$ not be $O(1)$. We also discard parameter sets where the mass spectra do not satisfy $10^{-3} \le m_\mu^2/m_\tau^2 \le 10^{-2}$ and $10^{-8} \le m_e^2/m_\tau^2 \le 10^{-6}$. Only parameter sets that satisfy both conditions are included in $\mathcal{C}_2$.

Second, we notice that parameter sets that satisfy the conditions tend to be concentrated around very small $\text{sup} \{ v/v_T, v_5/v_T \}$. Since we want to study $\theta_{13}$ over a large range of $\epsilon$, we generate $10,000$ parameter sets (satisfying the conditions) for $v_T$ a log flat random variable between $10^{16}$ and $10^{19} \, \text{GeV}$, $6,000$ sets for $v_T$ between $10^{16}$ and $10^{18} \, \text{GeV}$ and $4,000$ sets for $v_T$ between $10^{16}$ and $10^{17.5} \, \text{GeV}$. This ensures that the combined $20,000$ sets in $\mathcal{C}_2$ span a useful range of $\text{sup} \{ v/v_T, v_5/v_T \}$ that we can work with.

Finally, we explain the motivation behind Eq.~(\ref{eqn:yukawa-constraints1}) and (\ref{eqn:yukawa-constraints2}). From Eq.~(\ref{eqn:factorization2}), the zeroth-order term of $M_l$ is $\frac{v_H}{\Lambda} \left( A - \frac{y^T_m}{y^{T\prime}_m} C \right)$. This has eigenvalues
\begin{equation}
m_a =\sqrt{3} \frac{v_H}{\Lambda} \left| y_e v - \frac{y^T_m}{y^{T\prime}_m} y'_e v\right\vert, \quad
m_b, m_c = \frac{v_H}{\Lambda} \left\vert \sqrt{3} \left( y_m - \frac{y^T_m}{y^{T\prime}_m} y'_m \right) v \pm 3i \left( y^5_m - \frac{y^T_m}{y^{T\prime}_m} y^{5\prime}_m \right) v_5 \right\vert. \label{eqn:mass-eigenvalues-2}
\end{equation}
Eq.~(\ref{eqn:yukawa-constraints1}) and (\ref{eqn:yukawa-constraints2}) hence ensures that the zeroth-order eigenvalues satisfy the mass ratio conditions. Nonetheless, we note that only a very small fraction of random parameter sets generated this way end up being included in $\mathcal{C}_2$. The reason is that subleading corrections to the small eigenvalues from block diagonalisation can be much larger than the zeroth-order small eigenvalues themselves, especially for larger values of $\text{sup} \{ v/v_T, v_5/v_T \}$. As a result, the mass ratio conditions may be violated.

\subsection{$\mathcal{C}_3$, $\mathcal{C}_4$, $\mathcal{C}_5$ and $\mathcal{C}_6$} \label{sect:random2}

The collections $\mathcal{C}_3$, $\mathcal{C}_4$, $\mathcal{C}_5$ and $\mathcal{C}_6$ are used in Figure~\ref{fig:second-method}. The random parameters of interest are the VEVs $v$ and $v_5$, the Yukawas $y_e$, $y_m$ and $y_m^5$, and the changes $\delta a$ and $\delta b$ from the original values of $a$ and $b$. In all four collections, the VEVs are uniform random variables between $10^{15}$ and $10^{16} \, \text{GeV}$. For $\delta a$ and $\delta b$, we first simulate the deviation angle $\chi$ as log flat random variable between $10^{-4}$ and $10^{-1}$. Since $\chi^2 \simeq (\delta a)^2 + 2 (\delta b)^2/3$, we simulate $\delta a$ as a uniform random variable between $-\chi$ and $\chi$, and then derive $\delta b$ using $\delta b = \pm \sqrt{3(\chi^2 - (\delta a)^2)/2}$, with the signs randomly generated.

The difference between the four collections lie in the Yukawas, since the size of Yukawas is directly related to the size of the mass eigenvalues. For $\mathcal{C}_3$, all Yukawas are simply $O(1)$ uniform random complex variables, with real and imaginary parts between $-3$ and $3$. For $\mathcal{C}_4$, we generate $y_e$ and $y_m$ as $O(\frac{1}{1000})$ and $O(1)$ random complex variables. We also first generate a $O(0.1)$ random complex variable $\alpha$, from which $y_m^5$ is derived using the relation
\begin{equation}
y_m^5 = -\frac{i}{\sqrt{3}} \frac{v}{v_5} y_m (1-\alpha)
\end{equation}
These choices are made to increase the likelihood of the eigenvalues satisfying the correct mass ratio. For $\mathcal{C}_5$ and $\mathcal{C}_6$, the procedures are similar to that of $\mathcal{C}_4$, except that $y_e$ and $\alpha$ are further reduced to increase the likelihood of satisfying the (unphysical) smaller mass ratios.

\bibliography{A4-project-v3}
\bibliographystyle{apsrev4-1}
\end{document}